\documentclass[twocolumn,showpacs,aip,reprint,amsmath,amssymb]{revtex4-1}
\usepackage{graphicx}
\usepackage[colorinlistoftodos]{todonotes}
\usepackage{multirow}

\newcommand{\vect}[1]{{\bf {#1}}}
\newcommand{\mat}[1]{{\underline{\underline{#1}}}}
\newcommand{\vc}[1]{{\underline{#1}}}
\newcommand{\fnr}{{\left(\vect{r}\right)}}
\newcommand{\fnk}{{\left(\vect{k}\right)}}

\newcommand{\fnint}{{\int\mathcal{D}}}
\newcommand{\rint}{{\int d\vect{r}\,}}

\begin{document}


\title{A multi-species exchange model for fully fluctuating polymer field theory simulations}

\author{Dominik D\"uchs}
\thanks{These authors contributed equally to this work.}
\author{Kris T.\ Delaney}
\email[Electronic address: ]{kdelaney@mrl.ucsb.edu}
\thanks{These authors contributed equally to this work.}
\affiliation{Materials Research Laboratory, University of California, Santa Barbara, CA 93106, USA} 
\author{Glenn H.\ Fredrickson}
\email[Electronic address: ]{ghf@mrl.ucsb.edu}
\affiliation{Materials Research Laboratory, University of California, Santa Barbara, CA 93106, USA} 
\affiliation{Departments of Materials and Chemical Engineering, University of California, Santa Barbara, CA 93106, USA} 

\date{\today}

\begin{abstract}
Field-theoretic models have been used extensively to study the phase behavior of
inhomogeneous polymer melts and solutions, both in self-consistent mean-field
calculations and in numerical simulations of the full theory capturing composition
fluctuations.
The models commonly used can be grouped into two 
categories, namely {\it species} models and {\it exchange} models. 
Species models
involve integrations of functionals that explicitly depend on fields originating both from 
species density operators and their conjugate chemical potential fields. 
In contrast,
exchange models retain only linear combinations of the chemical potential fields.
In the two-component case, development of exchange models has been instrumental in enabling 
stable complex Langevin (CL) simulations of the full complex-valued theory.
No comparable stable CL approach has yet been established for field theories of the species type. 
Here we introduce an extension of the exchange model to an arbitrary number of components, namely the
multi-species exchange (MSE) model, which greatly expands the classes of soft
material systems that can accessed by the complex Langevin simulation technique.  
We demonstrate the stability and accuracy of the MSE-CL sampling approach using numerical simulations 
of triblock and tetrablock terpolymer melts, and tetrablock quaterpolymer melts.
This method should enable studies of a wide range of fluctuation phenomena in multiblock/multi-species
polymer blends and composites.
\end{abstract}

\pacs{}

\maketitle

\section{Introduction}
Self-consistent field theory (SCFT) is a versatile and well-established
numerical method for studying the mesoscopic self-assembly and equilibrium thermodynamic behavior of polymeric melts and
solutions\cite{fredrickson06,schmid98}. 
In contrast to microscopic simulation methods like molecular
dynamics\cite{frenkel} or coarse-grained Monte Carlo\cite{landau00}, partition functions containing
particle-particle interactions are transformed mathematically such that the
fundamental objects in SCFT are auxiliary chemical potential fields and their conjugate
monomer species densities. 
While the field theories underpinning SCFT are formulated in continuous space, 
a range of numerical methods are available including 
projection of the fields onto a finite set of basis functions\cite{matsen94,matsen03}, 
real-space solutions via finite differences on a computational lattice\cite{fleer79}, 
or pseudo-spectral solutions on a collocation grid invoking fast Fourier Transforms (FFTs)\cite{rasmussen02,tzeremes02}. 
In all of these techniques the continuum SCFT equations are transformed into a large but finite 
set of non-linear equations that are solved iteratively by relaxation methods. 
The stationary solutions are discrete representations
of the saddle-points of the field theory, and hence constitute \emph{mean-field approximations} to 
the full theory. 
For each iteration of the field relaxation, a modified diffusion equation must be solved to deduce the 
statistical properties of non-interacting chains in a set of static external chemical-potential fields; this
constitutes the vast majority of the computational cost of the method.

For inhomogeneous polymer systems, a frequent topic of study is the self-assembly of constituents
into a variety of complex mesoscale morphologies (e.g., lamellae, bicontinuous networks, 
or hexagonally packed cylinders).
SCFT allows phase diagrams to be constructed in the space of architectural
parameters (e.g., the block molecular weights in copolymers) and segmental interaction
parameters between the constituent species. 
Being a mean-field approximation, however,
SCFT neglects fluctuations in composition profiles; this approximation is 
valid only for very dense systems, such as melts, in regions of the phase diagram that 
are sufficiently removed from critical phase transitions.
To capture fluctuation effects, simulations of the full field theory are needed.

A major impediment to a straightforward implementation of simulations of the full theory
is what is known in the quantum field-theory literature as the {\it sign problem}: the occurrence
of rapidly oscillating complex exponential factors making the direct evaluation of
statistical averages intractable. 
Until recently, strategies for handling the sign problem in polymer field theories
had been developed and applied only for two-species systems.
In one attempt the {\it partial saddle-point method} approximates the full theory by
restricting fluctuations to fields without rapid oscillations\cite{duechs03,stasiak/matsen:2013}.
This approach is suitable mainly for two-species incompressible systems as it neglects 
fluctuations in the pressure mode governing total local densities, but it 
nevertheless constitutes an uncontrolled approximation.
A more comprehensive solution, which captures fluctuations of all fields in the 
full theory, involves applying the complex Langevin (CL) method\cite{parisi83,klauder83} 
for generating importance-sampled sequences of field configurations. 
This method tolerates the sign problem by analytically continuing the theory into the space of 
complex fields, and sampling stochastically along high-dimensional paths that are locally of 
near-constant phase.
Estimates for physical properties (thermal averages) remain real valued.
This method has previously been used to study a variety of problems including the
fluctuation-induced shift in the order-disorder transition in symmetric\cite{ganesan01} and asymmetric\cite{lennon08_2} 
diblock copolymers, and the complexation of oppositely charged polyelectrolytes\cite{lee08}.

Not surprisingly, a simulation method capturing the full statistical field
theory faces several numerical challenges. 
As in SCFT, diffusion equations must be solved at each CL field iteration step. 
In contrast to SCFT however, where the purpose of iterating the fields is to approach a
saddle-point solution, field updates in CL serve to sample representative field 
configurations with probability consistent with the partition function.
An important efficiency consideration then is to ensure that field updates are performed 
in a manner that is both stable and accurate, and to propagate the system 
as quickly as possible between statistically uncorrelated field configurations.
Lennon \emph{et al}.\ have discussed the performance of several explicit and
implicit field-update algorithms in selected two-component test cases\cite{lennon08}.
Another numerical concern is the choice of a suitable algorithm to solve the diffusion
equations in the presence of rough, stochastically generated fields. 
Recently Audus \emph{et al}.\ demonstrated that higher-order diffusion solvers
can be less stable than low-order solvers when subjected to fields containing white noise\cite{audus13}.
While these considerations are important from a practical standpoint, a parallel concern is 
the stability of the CL equations in multi-component systems, which are of growing interest
in the design and fabrication of novel nanostructured materials\cite{BatesScience2012}. 
In particular: \emph{How can one construct a field-theoretic model and a corresponding CL 
stochastic dynamics for sampling the fields that is stable for systems containing more than two monomer species?} 
We address this challenge in the present article by generalizing the commonly used
exchange model to polymer systems with more than two species, demonstrating for the first
time that CL can be applied to polymeric fluids with many chemically distinct components; 
the method in principle extends to an arbitrary number of species.

A detailed discussion of field-theoretic models for polymeric melts and
solutions can be found, e.g., in Ref.\ \onlinecite{fredrickson06}. 
We here briefly review the two broad classes of models, which are in principle thermodynamically equivalent.
In density-explicit species-type models, a pair of thermodynamically conjugate 
complex fields arises \emph{for each monomer/segment species}: an auxiliary chemical-potential field and a
density field.
In addition, incompressible models feature a ``pressure'' field conjugate to the total
density at each point in space.
Hence, for $S$ species, a typical density-explicit species model formally involves $2S+1$ independent fields.
In contrast, \emph{exchange-type models}, which prior to this work 
have been formulated only for $S=2$ component systems, decouple the interactions
using the normal modes of the interaction matrix rather than the species fields themselves;
the decoupling allows redundant degrees of freedom to be integrated out of the problem
(provided the pair interaction is invertible).
%
%
As a result, exchange-type models utilize only $S$ fields for $S$ components, and
are therefore more efficient in their representation of a multi-component system.
This is a first indication that exchange models are better suited to CL than species-type models,
as imposing independent fluctuations on an over-complete set of fields can be imagined to
produce numerical instabilities.
Indeed, even for a simple two-species system, such as an incompressible AB diblock copolymer melt, 
one can show that the linearized CL equations for the species model possess
eigenvalues of mixed sign, symptomatic of an unstable scheme.
In contrast, the exchange formulation of the model has been shown to enable stable CL simulations of the
same diblock copolymer melt model\cite{ganesan01,lennon08,lennon08_2}.
We note that a third type of simulation method has been recently explored\cite{Koski/Huikuan/Riggleman:2013}, 
utilizing an exchange mapping for \emph{each pair} of species.
The method was demonstrated to be stable for conducting simulations of a nanocomposite melt of linear AB polymer 
chains interacting with nanoparticles of a third distinct species, though the method requires $2S+1$ fields for $S$ species.
We expect that the MSE approach presented here will be more efficient and stable as the number of distinct species
is increased.

This paper is organized as follows.
In Section II, we generalize the two-species exchange model to an arbitrary number of
species for incompressible and weakly compressible melt models.
In Section III we devise a complex Langevin (CL) sampling scheme for conducting 
field-theoretic simulations based on the new {\it multi-species exchange (MSE)} model, 
and discuss numerical aspects that affect efficiency. 
In section IV, we apply our model and methods to symmetric ABC triblock and 
ABCA$^\prime$ tetrablock terpolymer melts, and to ABDC tetrablock quaterpolymer melts; 
we show that CL sampling is stable 
both for disordered melts at low segregation strength and for ordered mesophases
at intermediate segregation strength. 
We further demonstrate that the sampling is accurate by comparing structure factors 
in the disordered melt to those obtained by the random phase approximation, and through
the recovery of the correct mean-field limit. 
Finally, we demonstrate that the compressible model fully recovers the 
incompressible one in the appropriate limit, while in some cases permitting more stable and efficient simulations.

\section{Theory}
\label{sec:theory}
\subsection{The multi-species exchange model of strictly incompressible melts}
\paragraph*{Coarse-grained particle model:}
We begin with the following canonical partition function for a coarse-grained particle model of a
multi-species polymer melt:
\begin{widetext}
\begin{equation}
  \mathcal{Z}_c = \frac{1}{\prod_{p=1}^{P}\lambda_T^{3 n_pN_p} n_p! }\prod_{i=1}^n\int\mathcal{D}\vect{r}_i e^{-\beta U_0 -\beta U_1}\delta\left[\sum_{j=1}^S\hat{\rho}_j\fnr -\rho_0\right].
  \label{eqn:ce_incmelt_particlepartfn}
\end{equation}
\end{widetext}
In a volume $V$, there are a total of $n$ molecules with $P$ distinct molecule types.
Each molecule type $p\in\left[1,P\right]$ has $n_p$ indistinguishable copies, such that $n=\sum_{p=1}^Pn_p$, and $N_p$ degrees of freedom.
Assuming the molecules are all linear polymer chains (without loss of generality on the field-theory transformation), 
the degree of polymerization of each chain type is $N_p$, and we choose arbitrarily a reference $N$
such that $N_p = N\alpha_p$.
The overall density of monomers in the melt is $\rho_0 = \sum_{p=1}^P n_p N_p/V$, and we assume that all statistical
segments have a common reference volume $v_0 = 1/\rho_0$.
$\lambda_T$ is the thermal wave length.
A total of $S$ distinct chemical (``monomer'') species are distributed in arbitrary sequences in the chains.
For a continuous-chain model, as employed throughout this work, the species density operators are
\begin{equation}
  \hat{\rho}_j\fnr  = \sum_{p=1}^{P} \sum_{i=1}^{n_p} \int_{s\in j}ds\, \delta\left(\vect{r}-\vect{r}^p_i\left(s\right)\right),
\end{equation}
where the integration bound $s\in j$ includes all statistical segments of chemical species $j$ in the chain-contour integral
(i.e., repeated blocks, e.g., in linear ABA polymers, are handled implicitly by repeated summation).
$\vect{r}_i^p\left(s\right)$ is the space-curve of polymer chain $i$ of type $p$, parameterized as a continuous linear 
filament with contour variable $s$.
The functional integrals $\mathcal{D}\vect{r}_i$ run over the
possible space curves $\vect{r}\left(s\right)$ for the $i$th polymer chain.
The delta functional in Eqn.~\ref{eqn:ce_incmelt_particlepartfn} strictly enforces
incompressibility in the melt by 
projecting onto a configuration subspace for which the microscopic densities
sum to a uniform constant segment density ($\rho_0$) throughout space.

The energy weights in Eqn.~\ref{eqn:ce_incmelt_particlepartfn} are $\beta U_0$, 
the conformational statistics of a single chain in free space (e.g., chain stretching term such as the Wiener weight
for a continuous Gaussian chain), and $\beta U_1$, the potential energy for 
interactions between distinct segments.
For linear continuous chains with Gaussian stretching statistics, the conformation term is
\begin{equation}
  \beta U_0 =  \sum_{p=1}^{P}\sum_{i=1}^{n_p} \int_0^{N\alpha_p}ds\, \frac{3}{2b\left(s\right)^2}\left|\frac{d\vect{r}^p_i\left(s\right)}{ds}\right|^2,
\end{equation}
where $b\left(s\right)$ is the statistical segment length of a monomer at contour position $s$.


For Flory-like contact interactions, the potential energy term is
\begin{eqnarray}
  \beta U_1 & = & \frac{1}{2\rho_0 N}\sum_{i,j=1}^S\rint  \hat{\rho}_i\fnr \chi_{ij}N\hat{\rho}_j\fnr \,\\
            & = & \frac{1}{2\rho_0 N}\rint  \vc{\rho}^T\mat{\chi}N\vc{\rho},
\end{eqnarray}
where $\vc{\rho}^T = \left(\hat{\rho}_1,\ldots,\hat{\rho}_S\right)$ and $\mat{\chi} = \left(\chi_{ij}\right)$ with $i,j\in\left\{1,\ldots,S\right\}$.
In this model it is assumed that $\chi_{ii}=0 \enskip \forall i \in \left\{1,\ldots,S\right\}$ and $\chi_{ij} = \chi_{ji} \enskip \forall i,j \in \left\{1,\ldots,S\right\}$.
$\beta U_1$ is responsible for the many-body character of the problem, and will be replaced by particle-field interactions
using Hubbard-Stratonovich transformations. However, since the interaction matrix $\chi_{ij}N$ is usually not positive definite, 
direct Hubbard-Stratonovich transformation is usually not possible, and must instead be accompanied by decomposition into
normal modes (``exchange mapping'').

\paragraph*{Field Theory:}
Appendix \ref{sec:appendix_incomp} details the derivation of an exchange-mapped field 
theory corresponding to the canonical partition function
Eqn.~\ref{eqn:ce_incmelt_particlepartfn}.
The resulting partition function is
\begin{equation}
  \label{eqn:Z_incomp_mse_body}
  \mathcal{Z}_c = \mathcal{Z}_0 \fnint \mu_1\ldots\fnint \mu_{S-1}\fnint \mu_+\, e^{-H\left[\left\{\mu_i\right\},\mu_+\right]}\\
\end{equation}
with
\begin{widetext}
\begin{equation}
  \label{eqn:H_incomp_mse_body}
  H\left[\left\{\mu_i\right\},\mu_+\right] = C\left[-\sum_{i=1}^{S-1}\frac{1}{2d_i}\rint  \mu_i^2\fnr  + \sum_{i,j=1}^{S-1}\frac{O_{ji}\chi_{jS}N}{d_i}  \rint \mu_i\fnr - \rint \mu_+\fnr  - \sum_{p=1}^P \frac{V\phi_p}{\alpha_p} \ln Q_p\left[\mat{A}\vc{\mu}\right]\right],
\end{equation}
\end{widetext}
and $\mathcal{Z}_0$ is a constant, including ideal gas terms, fully defined in Appendix~\ref{sec:appendix_incomp}.
Eqns.\ \ref{eqn:Z_incomp_mse_body} and \ref{eqn:H_incomp_mse_body} constitute a full description of the field theoretic canonical partition function.
The fields $\mu_1,\ldots,\mu_{S-1},\mu_+$ are auxiliary \emph{exchange}-mapped chemical potential fields that
are related to the \emph{species} chemical potential fields $\psi_A\fnr N,\ldots,\psi_S\fnr N$ by a linear transformation.
$\mu_+$ is responsible for enforcing incompressibility of the melt.
$d_i$ and $O_{ij}$ are respectively the eigenvalues and eigenvectors of the $\left(S-1\right)\times\left(S-1\right)$ matrix 
$X_{ij} = \chi_{ij}N-2\chi_{iS}N$.
All lengths are non-dimensionalized using the radius of gyration of a reference ideal polymer chain, $R_g=b\left(N/6\right)^{1/2}$, with reference statistical segment
length $b$ and reference polymerization degree $N$. $C=\rho_0 R_g^3/N$ is a dimensionless polymer chain number density parameter. 
$Q_p$ is the partition function of a single molecule of type $p$ experiencing the $S$ $\mu$ fields, while $\phi_p$ is the overall volume fraction 
of the system occupied by molecules of type $p$ (i.e., $\sum_p \phi_p = 1$).
The functional integrals $\int\mathcal{D}\mu_i$ are over all configurations of real fields if $d_i<0$, and over purely imaginary fields 
otherwise (i.e., fields corresponding to repulsive interactions, which have saddle point configurations of physical relevance that are purely imaginary,
have been Wick rotated to bring the saddle-point onto the real axis, as discussed in Appendix \ref{sec:appendix_incomp}).
The $S\times S$ matrix $\mat{A}$ specifies the linear transformation from exchange-mapped fields to the species chemical 
potential fields, i.e., $\psi_i\fnr N = \sum_j A_{ij}\mu_j\fnr$, which enter the single-chain partition function:
\begin{equation} 
  A_{ij} =
\begin{pmatrix} O_{ij} & 1 \\  0 & 1 \end{pmatrix}.
\label{A}
\end{equation}
The inverse transformation $\mat{A}^{-1}$ takes the form
\begin{equation}
  A^{-1}_{ij} = \left(\begin{array}{cc}
  O_{ji} & X_i\\
  0 & 1
  \end{array}\right),
\end{equation}
where $X_i = -\sum_{k=1}^{S-1}O_{ki}$.
The need for the inverse transformation arises only during initialization of 
$\left\{\mu_i\right\}, \mu_+$ based on stored species-field configurations $\left\{\psi_i N\right\}$.
It is usually more convenient to initialize calculations from species fields because the 
precise meaning of exchange fields can change significantly with system parameters.
Once initialized, however, a CL or SCFT simulation will proceed without requiring the inverse transformation.

The single-chain partition functions $Q_p\left[\mat{A}\vc{\mu}\right]$ are evaluated for continuous linear chains according to
\begin{equation}
  Q_p = \frac{1}{V} \rint q_p\left(\vect{r}, \alpha_p; \left[\mat{A}\vc{\mu}\right]\right),
\label{Qp}
\end{equation}
where the propagators $q_p(\vect{r}, s)$ are governed by the diffusion equation
\begin{eqnarray}
  \frac{\partial}{\partial s} q_p\left(\vect{r},s\right) &=& \left[\frac{b\left(s\right)^2}{b^2}\nabla^2-N\psi_p\left(\vect{r},s\right)\right] q_p\left(\vect{r},s\right),\\
  q_p\left(\vect{r},0\right) &=& 1,
\label{diff_forward}
\end{eqnarray}
which must be solved along the contour variable $s\in\left[0,\alpha_p\right]$.
The external field $N\psi_p\left(\vect{r}, s\right)$ is equal to
$\psi_{A}\fnr N$, $\psi_{B}\fnr N$, etc., depending on the chemical identity of 
the segment at contour position $s$, where $\psi_i\fnr N = \sum_j A_{ij} \mu_j\fnr$ are
the ``species'' chemical potential fields.
The forward propagator, $q_p(\vect{r},s)$, can be combined with
a backward propagator, $q_p^{\dagger}(\vect{r},s)$, governed by
\begin{eqnarray}
  \frac{\partial}{\partial s} q_p^{\dagger}\left(\vect{r},s\right) &=& \left[\frac{b\left(s\right)^2}{b^2}\nabla^2-N\psi_p^\dagger\left(\vect{r},s\right)\right] q_p^{\dagger}\left(\vect{r}, s\right),\\
  q_p^{\dagger}\left(\vect{r},0\right) &=& 1,
\label{diff_backward}
\end{eqnarray}
with $\psi_p^{\dagger}$ defined by traversing polymer $p$ in the contour parameter $s$ from the opposite chain end,
to deduce dimensionless monomeric density operators, $\varphi_i\left(\vect{r}\right)$, which are required for 
CL and mean-field relaxation dynamics.
The density operators are
\begin{equation}
\label{rho_i}
\varphi_j(\vect{r}; [\{\mu_{i}\},\mu_{+}] ) = \sum_p \frac{\phi_p}{Q_p \alpha_p} \int_{s\in j} ds \, q_p(\vect{r},s) q_p^{\dagger}(\vect{r}, s),
\end{equation}
where $s\in j$ denotes contour segments along chain $p$ occupied by monomers of species type $j$. These 
segment densities are normalized so that their expectation values over all field configurations weighted 
by $\exp (-H)$ yield the \emph{local} volume fraction of type-$j$ segments, i.e., 
$\rho_0\left<\varphi_j\fnr\right>_{\left\{\mu\right\}} = \left<\hat{\rho}_j\right>_{\left\{\vect{r}_i^p\right\}}$, where the former
is a field-based operator and the latter is a microscopic density operator averaged over particle coordinates. It is straightforward to extend this 
definition of propagators and density operators to non-linear/branched polymer architectures, and to discrete-chain models.

Although not used in the present study, we note for completeness the form of the action in the grand canonical ensemble:
\begin{widetext}
\begin{equation}
H_G\left[\left\{\mu_i\right\},\mu_+\right] = C\left[-\sum_{i=1}^{S-1}\frac{1}{2d_i}\rint  \mu_i^2\fnr  + \sum_{i,j=1}^{S-1}\frac{O_{ji}\chi_{jS}N}{d_i}  \rint \mu_i\fnr - \rint \mu_+\fnr\right]  - \sum_{p=1}^P z_p V \ln Q_p\left[\mat{A}\vc{\mu}\right],
\end{equation}
\end{widetext}
where $z_p$ is the activity of chain $p$, related to the chemical potential $\mu_p$ by $z_p = \exp\left(\beta \mu_p\right)$.
For a strictly incompressible melt, the compositional redundancy manifests as an arbitrary scaling of chain activities: only ratios of $z_p$ 
affect the thermodynamics.
All other symbols have the same definitions as in the canonical ensemble.

\paragraph*{Complex Langevin Sampling:}
It is now appropriate to discuss the use of the {\it MSE} model for complex Langevin (CL) simulations of
polymer systems containing an arbitrary number of chains $P$ and species $S$. 
Key to devising a stable CL scheme is writing Langevin equations not for the species potentials, but for the 
generalized eigenmode potentials $\left\{\mu_j\right\}, \mu_+$.
In the canonical ``diagonal''  CL dynamics adopted here, the relaxational part of the field dynamics
is in the direction of the generalized force acting on that same field.  
Including random force terms, the full stochastic CL equation for fields $\left\{\mu_j\right\}, \mu_+$ are:
\begin{equation}
\label{eqn:CL}
  \frac{\partial\mu_{i}\left(\vect{r}, t\right)}{\partial t} = -\lambda_{i}\gamma_{i}^2\frac{\delta H\left[\left\{\mu_{i}\right\}\right]}{\delta \mu_{i}\left(\vect{r},t\right)}
  + \gamma_{i} \eta_{i}\left(\vect{r},t\right),
\end{equation}
where $t$ is a fictitious simulation time variable, $\lambda_{i}>0$ is a real-valued relaxation rate parameter, and $i$ is
an $S$-dimensional index spanning $\left[1,S-1\right]$ and $+$. 
The parameters $\gamma_{i}$ arise from the Wick rotation of the pressure and pressure-like fields, defined by
\begin{eqnarray}
  \gamma_{+} &=&  i, \\
  \gamma_{j} &=& \left\lbrace{\begin{array}{c}
1, \, \;  d_{j} < 0 \\
i, \, \;  d_{j} > 0
\end{array}} \right.,
\label{gamma}
\end{eqnarray}
where $i=\sqrt{-1}$. 
Finally, the fields $\eta_{i}({\bf r}, t)$ are
purely real and independent Gaussian white (in both space and time) noise
terms obeying the fluctuation dissipation theorem:
\begin{eqnarray}
  \left<\eta_{i}\left(\vect{r},t\right)\right> &=& 0 \nonumber \\
  \left<\eta_{i}\left(\vect{r},t\right)\eta_{i}\left(\vect{r}^\prime,t^\prime\right) \right> &=& 2 \lambda_{i} \delta\left(\vect{r}-\vect{r}^\prime\right)\delta\left(t-t^\prime\right).
\label{fluc_diss}
\end{eqnarray}

The CL equations (Eqns.\ \ref{eqn:CL}) involve thermodynamic forces that are expressed 
as functional derivatives of the action with respect to the generalized field 
variables.  The relevant derivatives can be expressed in terms of the generalized 
potentials and the species density operators by the following expressions:
\begin{eqnarray}
  \frac{\delta H\left[\left\{\mu_i\right\},\mu_+\right]}{\delta \mu_{i} ({\bf r})} &=&
  \frac{C}{d_i}\left[-\mu_{j}\left(\vect{r}\right) + \sum_{j=1}^{S-1}O_{ji}\chi_{jS}N\right]\nonumber\\
  & +& C\sum_{j=1}^{S-1} O_{ji}\varphi_{j}\left(\vect{r}\right)\\
  \frac{\delta H\left[\left\{\mu_j\right\},\mu_+\right]}{\delta \mu_{+} \left(\vect{r}\right)} &=&
  C\left[\sum_{j=1}^{S}\varphi_{j}({\bf r}) - 1\right]
  \label{eqn:fplus}
\end{eqnarray}

An interesting observation from Eqn.\ \ref{eqn:fplus} is that the force on the $\mu_+$ field has
no explicitly linear component. 
Furthermore, the spatial average of the sum of normalized species density operators is guaranteed to be 
unity for \emph{any} configuration of the fields $\left\{\mu_i\right\},\mu_+$.  
Hence, the spatial average of the force on $\mu_+$ is zero for \emph{any} $\mu_+$ configuration. 
This reflects an invariance of the species densities to a shift in the average value of $\mu_+$, which 
originates from the invariance of the exponential form of the incompressibility delta functional to such a shift. 
A practical consequence is that upon application of noise, the average value of the $\mu_+$ field will drift arbitrarily.
However, all physical properties should be invariant to this drift.
This invariance can be seen in $H$, because the linear term in $-n\ln Q_p$ (Sec.\ \ref{sec:wie}) exactly cancels the explicit $\mu_+$ integral in $H$.
We have nevertheless found it convenient to constrain $\rint \mu_+\fnr=0$ during our CL and SCFT simulations, a choice
with no thermodynamic consequences.

An interesting feature of the complex Langevin method is that instantaneous samples of thermodynamic operators 
are complex. 
However, as required on physical grounds, all observables become real upon thermal (time) averaging.

We note that eliminating the noise terms in Eqns.\ \ref{eqn:CL} leads to deterministic relaxation equations
that have mean-field saddle-point solutions as the long-time stationary state; i.e., all fields relax until
\begin{equation}
  0 = \left.\frac{\delta H\left[\left\{\mu_i\right\},\mu_+\right]}{\delta \mu_{i,+}\fnr}\right|_{\left\{\mu_i\right\}=\left\{\mu_i^\star\right\}, \mu_+=\mu_+^\star}
\end{equation}
is satisfied.
One convenient aspect of SCFT is the immediate availability of the Helmholtz free 
energy $\beta F = -\ln \mathcal{Z}_c \approx -\ln \mathcal{Z}_0 + H^\star$, since under the saddle-point
approximation $\mathcal{Z}_c \approx \mathcal{Z}_0 \exp\left(-H\left[\left\{\mu_i^\star\right\},\mu_+^\star\right]\right)$.
This feature makes constructing SCFT phase diagrams straightforward for a predetermined set of candidate phases.

Finally, it is important to recognize that all saddle-point field configurations are independent of the $C$ chain number density
in this model.
This invariance means that the molecular weight of polymers in the melt enters only in unison with the interaction
parameters through the $\chi N$ terms.
In contrast, inclusion of composition fluctuations generates a family of models for different $C$ chain densities, where
the $C$ parameter depends implicitly on the number of statistical segments present in the polymer chains.
Hence, beyond-mean-field simulations acquire a molecular-weight dependence beyond $\chi N$ to all thermodynamic observables.

\subsection{Weakly compressible multi-species melt}
\paragraph*{Coarse-grained particle model:}
In this case, the following particle model is used:
\begin{widetext}
\begin{equation}
  \mathcal{Z}_c = \frac{1}{\prod_{p=1}^{P} \lambda_T^{3 n_pN_p}n_p! }\int\ldots\int\prod_{i=1}^{n}\mathcal{D}\vect{r}_i \exp\left(-\beta U_0 -\beta U_1 - \beta U_2\right).
  \label{eqn:ce_hcmelt_particlepartfn}
\end{equation}
\end{widetext}
$\beta U_0$ and $\beta U_1$ are identical to the previous model, 
while the delta functional that enforced incompressibility has been replaced by a
Helfand\cite{Helfand75} weak compressibility penalty $\beta U_2 = \frac{\zeta N}{2\rho_0N}\rint  \left(\sum_{j=1}^S\hat{\rho}_j\fnr -\rho_0\right)^2$.
The Helfand compressibility ensures that the thermally and spatially averaged density is $\rho_0$ (mass conservation), but local fluctuations around the
spatial average density are permitted in the thermally averaged density profiles.
This type of model can be numerically advantageous over the strictly incompressible
model by making the complex Langevin dynamics less stiff, at the expense of introducing one extra parameter ($\zeta$).

\paragraph*{Field Theory:}
Appendix \ref{sec:appendix_comp} details the derivation of the exchange-mapped field theory for the compressible melt canonical
partition function (Eqn.\ \ref{eqn:ce_hcmelt_particlepartfn}).
The resulting partition function is
\begin{equation}
  \label{eqn:Z_comp_mse_body}
  \mathcal{Z}_c = \mathcal{Z}_0 \fnint \mu_1\ldots\fnint\mu_S \,e^{-H\left[\left\{\mu_i\right\}\right]},
\end{equation}
with action
\begin{widetext}
\begin{equation}
  H\left[\left\{\mu_i\right\}\right]  = C\left[-\sum_{i=1}^S \frac{\left(\zeta N\right)^{-\frac{1}{2}}}{2d_i}\rint \mu_i^2\fnr - \sum_{i,j=1}^S\frac{O_{ji}}{d_i\left(\zeta N\right)^{-\frac{1}{2}}}\rint\mu_i\fnr - V\sum_{p=1}^P \frac{\phi_p}{\alpha_p}\ln Q_p\left[\mat{O}\vc{\mu}\right]\right],
\end{equation}
\end{widetext}
$\mathcal{Z}_0$ a constant, and $d_i$ and $O_{ij}$ are eigenvalues and
eigenvectors of the $S\times S$ matrix
$\left(\mat{\chi}N\left(\zeta N\right)^{-\frac{1}{2}} + \left(\zeta N\right)^\frac{1}{2}\mat{\openone}\right)$,
where $\mat{\openone}$ is an $S\times S$ matrix with all entries equal to $1$.
Other symbols are as defined in the previous section.
Note that all sums include a full $S$ count of exchange-mapped fields, and there is no separate
pressure field.
The transformation matrices from exchange to species fields and vice versa
are directly $\mat{O}$ and $\mat{O}^T$ respectively.
For reference, the action of the grand partition function in this model is
\begin{widetext}
\begin{equation}
H_G\left[\left\{\mu_i\right\}\right]  = C\left[-\sum_{i=1}^S \frac{\left(\zeta N\right)^{-\frac{1}{2}}}{2d_i}\rint \mu_i^2\fnr - \sum_{i,j=1}^S\frac{O_{ji}}{d_i\left(\zeta N\right)^{-\frac{1}{2}}}\rint\mu_i\fnr\right] - V\sum_{p=1}^P z_p Q_p\left[\mat{O}\vc{\mu}\right].
\end{equation}
\end{widetext}
In contrast with the incompressible case, the absolute magnitude of chain activities $\left\{z_p\right\}$ is not
arbitrary; the total mass of the system is not specified by parameter constraints, and
the absolute values of $z_p$ combine with the compressibility parameter to determine the equilibrium
monomer density of the system.
In this case, the quantity $\rho_0$ should be considered to be the inverse of a reference volume of a statistical segment.

\paragraph*{Complex Langevin Sampling:}
We use the following equation of motion for all fields in the Helfand-type MSE model:
\begin{equation}
  \partial_t \mu_i\fnr = -\lambda_i\gamma_i^2\frac{\delta H}{\delta \mu_i\fnr} + \gamma_i\eta_i
  \label{eqn:comp_eom}
\end{equation}
where the forces are given by
\begin{eqnarray}
  \frac{\delta H}{\delta \mu_i\fnr} & = & C\left[-\frac{\left(\zeta N\right)^{-1/2}}{d_i}\mu_i\fnr -\frac{1}{d_i\left(\zeta N\right)^{-\frac{1}{2}}} \sum_{j=1}^SO_{ji}\right. \nonumber\\
                                    &+&\left.\sum_{j=1}^{S}O_{ji}\varphi_j\left(\vect{r}\right)\right].
  \label{eqn:comp_forces}
\end{eqnarray}

One point of caution is that the spectrum of eigenvalues becomes very broad if $\zeta N$ becomes large (approaching the incompressible limit), so that 
the relaxation rates of different fields are strongly disparate. 
Efficient sampling can often be restored by adjusting the $\lambda_i$ parameters appropriately.
An analysis of efficiency and mobility optimization is provided in Section \ref{sec:results}.

\section{Numerical Methods}
\label{sec:numerics}
Our numerical methods are based on pseudospectral collocation with periodic boundary conditions and second-order operator splitting for contour-stepping
the modified diffusion equations\cite{rasmussen02, tzeremes02, audus13, lennon08}, which are solved as an inner loop for each instantaneous configuration
of the auxiliary fields $\left\{\mu\right\}$.
Pseudospectral collocation with plane waves is ideally suited for complex Langevin simulations because it makes accessible 
large simulation cells (e.g., beyond the solution/melt correlation length) and does not rely on symmetric field configurations. 
Consequently, spatially decorrelated noise can be applied to the fields discretized on the computational lattice to affect unbiased stochastic sampling.

We find counter to intuition that solving the diffusion equation with a low-order (second) solver in the presence of 
rough fields makes for more stable CL trajectories\cite{audus13} when compared with nominally higher-order solvers.
In all of the present work, unless otherwise specified, we use second-order operator splitting with a contour step 
size of $\Delta s = 0.01N$ for computing the chain propagators (Eqns.\ \ref{diff_forward}, \ref{diff_backward}).
It now remains to discuss the relevant discretizations for the CL equations of motion, which stochastically evolve field configurations in an outer loop.

The complex Langevin (CL) method\cite{fredrickson06,lennon08,ganesan01,ghfreview02} 
allows sampling the full partition function (e.g., Eqns.\ \ref{eqn:Z_incomp_mse_body} and \ref{eqn:Z_comp_mse_body}) by moving along paths of approximate local constant phase,
thus avoiding adverse efficiency loss from strongly oscillating terms.
To numerically implement CL, the equations of motion must be discretized in space and time.
All of the CL equations of motion provided in Sections \ref{sec:theory}A and B have the generic continuum form
\begin{equation}
  \partial_t \mu\left(\vect{r},t\right) = -\lambda\gamma^2\frac{\delta H\left[\left\{\mu\right\}\right]}{\delta \mu\left(\vect{r},t\right)} + \gamma\eta\left(\vect{r},t\right), 
\end{equation}
with Gaussian white noise statistics $\left<\eta\left(\vect{r},t\right)\right>=0$, $\left<\eta\left(\vect{r},t\right)\eta\left(\vect{r}^\prime,t^\prime\right)\right> = 2\lambda\delta\left(\vect{r}-\vect{r}^\prime\right)\delta\left(t-t^\prime\right)$.
For spatial discretization, we first generate all functional derivatives in the continuum representation, then discretize the 
resulting functions onto the collocation mesh\footnote{A subtle issue should be noted here.
  The dimensions of a functional derivative $\delta H/\delta \mu\fnr$ are $\left[H\right]\left[\mu\right]^{-1}\left[V\right]^{-1}$. 
Hence, if the $H$ functional is discretized first, the partial derivative $\left(\Delta V\right)^{-1} \partial H\left(\vect{\mu}\right)/\partial \mu_i$ 
is the one that correctly approaches $\delta H/\delta \mu\fnr$ in the continuum limit.}. 
The resulting equation of motion on the computational lattice takes the form
\begin{equation}
  \partial_t \mu^\vect{r}\left(t\right) = -\lambda\gamma^2\left[\frac{\delta H\left[\left\{\mu\right\}\right]}{\delta \mu\left(\vect{r},t\right)}\right]^\vect{r} + \gamma\eta^\vect{r}\left(t\right), 
\end{equation}
now with noise statistics $\left<\eta^\vect{r}\left(t\right)\right>=0$, $\left<\eta^\vect{r}\left(t\right)\eta^{\vect{r}^\prime}\left(t^\prime\right)\right> = 2\lambda\left(\Delta V\right)^{-1}\delta_{\vect{r},\vect{r}^\prime}\delta\left(t-t^\prime\right)$,
where $\Delta V = V/M$, and $M$ is the number of cells on the spatial collocation mesh.
Notice that the equations of motion are local in space, which makes the computational effort of solving the CL field-update equations insignificant compared to the 
inner loop consisting of diffusion equations (Eqns. \ref{diff_forward}, \ref{diff_backward}).

\subsection{Complex Langevin time integrators}
We now discuss a variety of time integration algorithms with a range of computational costs and accuracy, stability, and efficiency profiles.
In practice, we have found that when used in tandem with appropriate 
semi-implicit time integration schemes, the discretized CL dynamics schemes presented in the previous section 
can be made both stable \emph{and} accurate for arbitrary numbers of species and choices of interaction parameters in the weak and intermediate
segregation regime. 
A number of the algorithms discussed here were presented in the context of two-species simulations by Lennon \emph{et al.}\cite{lennon08}, and 
all are compared in the context of multi-species simulations in Section \ref{sec:results}.
%

\paragraph*{Euler-Maruyama: }
The simplest discrete time stepper for evolving the complex Langevin equations is the Euler-Maruyama method, which is the stochastic
analog of forward Euler stepping.
The discretized equation of motion is
\begin{equation}
  \mu^{\vect{r},t+\Delta t} = \mu^{\vect{r},t}-\lambda\Delta t\gamma^2\left[\frac{\delta H\left[\left\{\mu\right\}\right]}{\delta \mu\left(\vect{r},t\right)}\right]^{\vect{r},t} + \gamma\eta^{\vect{r},t}, 
\end{equation}
with $\left<\eta^{\vect{r},t}\right>=0$, $\left<\eta^{\vect{r},t}\eta^{\vect{r}^\prime,t^\prime}\right> = 2\lambda\Delta t\left(\Delta V\right)^{-1}\delta_{\vect{r},\vect{r}^\prime}\delta_{t,t^\prime}$,
where $\Delta t$ is the time-step size.
This algorithm is accurate only to first order in the relaxation term and, being an explicit scheme, suffers from 
poor stability at very large time steps. 
However, the small time step required to control inaccuracies from the time-step bias 
mean that stability is less of a concern than accuracy when using this algorithm in CL simulations.
(SCFT, in contrast, is primarily limited by stability, because accurate trajectories are unimportant
provided the method eventually converges to a stationary saddle-point solution).

\paragraph*{1S first-order semi-implicit: } 
A stochastic extension to the SIS method introduced by Ceniceros and Fredrickson\cite{ceniceros04}, the 1S semi-implicit 
method takes the linear part of the relaxation term to the future time:
\begin{eqnarray}
  \mu^{\vect{r},t+\Delta t} = \mu^{\vect{r},t} & -&\lambda\Delta t\gamma^2\left(\left[\frac{\delta H\left[\left\{\mu\right\}\right]}{\delta \mu\left(\vect{r},t\right)}\right]^{\vect{r},t} + \left[\frac{\delta H\left[\left\{\mu\right\}\right]}{\delta \mu\left(\vect{r},t\right)}\right]_\mathrm{lin}^{\vect{r},t+\Delta t} \right.\nonumber\\
                                               &-&\left.\left[\frac{\delta H\left[\left\{\mu\right\}\right]}{\delta \mu\left(\vect{r},t\right)}\right]_\mathrm{lin}^{\vect{r},t}\right) + \gamma\eta^\vect{r}\left(t\right), 
\end{eqnarray}
with $\left<\eta^{\vect{r},t}\right>=0$, $\left<\eta^{\vect{r},t}\eta^{\vect{r}^\prime,t^\prime}\right> = 2\lambda\Delta t\left(\Delta V\right)^{-1}\delta_{\vect{r},\vect{r}^\prime}\delta_{t,t^\prime}$.
By using the linear response kernel for the homogeneous state, which is isotropic and translationally invariant (see Sections \ref{sec:numerics}B and C), the linearized part of the force can be written in the form 
\begin{equation}
  \left[\frac{\delta H\left[\left\{\mu\right\}\right]}{\delta \mu\left(\vect{r},t\right)}\right]_\mathrm{lin} \approx \left(\kappa\star\mu\right)\fnr
\end{equation}
with Fourier transform $\hat{\kappa}\fnk\hat{\mu}\fnk$, where $\left\{\vect{k}\right\}$ are vectors on the reciprocal lattice, and $\hat{\kappa}\fnk$ is a
linear-response kernel that can be computed analytically for any polymer model.
The 1S time stepper can then be rearranged to solve for $\mu^{t+\Delta t}$ in reciprocal space, yielding
\begin{equation}
  \hat{\mu}^{\vect{k},t+\Delta t} = \hat{\mu}^{\vect{k},t} - \frac{\Delta t \lambda \gamma^2}{1+\Delta t \lambda \gamma^2 \hat{\kappa}\left(\vect{k}\right)}\widehat{\left[\frac{\delta H\left[\left\{\mu\right\}\right]}{\delta \mu\left(\vect{r},t\right)}\right]}^{\vect{k},t} + \gamma\hat{\eta}^{\vect{k},t}
\end{equation}
The term preceding the Fourier transformed force is an effective $\vect{k}$-dependent time step 
that can be precomputed (and recomputed whenever the cell shape or volume changes).
Note that only force terms for which $\gamma^2 \hat{\kappa}\fnk>0$ for all $\vect{k}$
should be made semi-implicit in this fashion; other modes are destabilized by making this 
transformation.
In addition to linearized forces, explicitly linear terms in the force can be treated implicitly in the same way. 
All forces in the MSE scheme, except for $\delta H/\delta \mu_+$ in the strictly incompressible case, have explicitly linear terms that can be treated implicitly.

\paragraph*{Exponential time differencing:}
Exponential time differencing was introduced in the context of complex Langevin simulations by Villet and Fredrickson (Refs.\ \onlinecite{Villet2010, Villet2012}).
The method takes the linearized and explicitly linear parts of the relaxational force as an integrating factor over the interval $t\rightarrow t+\Delta t$, resulting
in
\begin{eqnarray}
  \mu^{\vect{k},t+\Delta t} = \mu^{\vect{k},t} &-&\left(\frac{1-e^{-\lambda\Delta t\gamma^2\hat{\kappa}\left(\vect{k}\right)}}{\hat{\kappa}\left(\vect{k}\right)}\right)\widehat{\left[\frac{\delta H\left[\left\{\mu\right\}\right]}{\delta \mu\left(\vect{r},t\right)}\right]}^{\vect{k},t} \nonumber\\
                                               &+&\left(\frac{1-e^{-\lambda\Delta t\gamma^2\hat{\kappa}\left(\vect{k}\right)}}{2\lambda\Delta t\gamma^2\hat{\kappa}\fnk}\right)\eta^{\vect{k},t},
\end{eqnarray}

\paragraph*{Weak second-order predictor-corrector methods:}
One of the disadvantages of the algorithms presented in the previous sections is the reliance on 
linearization of the relaxational terms of the CL dynamics to improve both stability and accuracy over each time step.
In practice, the linearization is performed with an isotropic, translationally invariant kernel 
that is accurate only for weak perturbations around a homogeneous state. 
In the limit of strongly segregated (inhomogeneous) polymer melts, the stability and accuracy of methods
employing such kernels is less certain. 
One option for moving beyond this limitation is to employ predictor-corrector algorithms, which divide 
the propagation of fields into two steps and average the force
between present and estimated future times for the corrected time step.
No assumption about the nature or amplitude of inhomogeneity is required.
These schemes offer weak second-order convergence to the continuous-time Langevin dynamics.

In the scheme due to \"Ottinger \cite{oettinger96} and Petersen\cite{petersen98} (PO), the
predictor step consists of an explicit forward step which, taken alone, reduces
to the Euler-Maruyama scheme.
The corrector step
then not only introduces a stabilizing implicit dependence but, in addition,
adds and subtracts the approximately linearized and explicitly linear derivatives at the future ($t + \Delta t$) and the predictor ($\tilde t$) 
time steps, respectively. 
The predictor and corrector update steps are:
\begin{eqnarray}
\label{PO}
\hat{\mu}^{\vect{k},\tilde t} = \hat{\mu}^{\vect{k},t} &-& \lambda\Delta t\gamma^2 \widehat{\left[\frac{\delta H\left[\left\{\mu\right\}\right]}{\delta \mu\left(\vect{r},t\right)}\right]}^{\vect{k},t} + \gamma\hat{\eta}^{\vect{k},t} \\
\hat{\mu}^{\vect{k},t+\Delta t} = \hat{\mu}^{\vect{k},t} & -&\frac{\lambda\Delta t\gamma^2}{2}\left(\widehat{\left[\frac{\delta H\left[\left\{\mu\right\}\right]}{\delta \mu\left(\vect{r},t\right)}\right]}^{\vect{k},t} +\widehat{\left[\frac{\delta H\left[\left\{\mu\right\}\right]}{\delta \mu\left(\vect{r},t\right)}\right]}^{\vect{k},\tilde{t}}  \right.\nonumber\\
                                                         &-&\left.\widehat{\left[\frac{\delta H\left[\left\{\mu\right\}\right]}{\delta \mu\left(\vect{r},t\right)}\right]}_\mathrm{lin}^{\vect{k},\tilde{t}} +\widehat{\left[\frac{\delta H\left[\left\{\mu\right\}\right]}{\delta \mu\left(\vect{r},t\right)}\right]}_\mathrm{lin}^{\vect{k},t+\Delta t}\right) \nonumber\\
                                                         &+& \gamma\hat{\eta}^{\vect{k},t},
\end{eqnarray}
Note that this method requires two full sets of solutions of the chain propagator modified diffusion equations to complete a time step, so
that the computational cost is doubled, but importantly
the \emph{same stochastic realization} of $\eta_{i}^{t}$ applies in both the predictor and corrector steps.

The PO method is significantly more accurate than the first order methods. 
If the linear response kernel is not employed in the corrector step, the method 
reduces to an Euler-Maruyama Predictor Corrector (EMPEC) method:
\begin{eqnarray}
\label{eqn:EMPEC}
\hat{\mu}^{\vect{k},\tilde t} = \hat{\mu}^{\vect{k},t} &-& \lambda\Delta t\gamma^2 \widehat{\left[\frac{\delta H\left[\left\{\mu\right\}\right]}{\delta \mu\left(\vect{r},t\right)}\right]}^{\vect{k},t} + \gamma\hat{\eta}^{\vect{k},t} \\
\hat{\mu}^{\vect{k},t+\Delta t} = \hat{\mu}^{\vect{k},t} & -&\frac{\lambda\Delta t\gamma^2}{2}\left(\widehat{\left[\frac{\delta H\left[\left\{\mu\right\}\right]}{\delta \mu\left(\vect{r},t\right)}\right]}^{\vect{k},t} +\widehat{\left[\frac{\delta H\left[\left\{\mu\right\}\right]}{\delta \mu\left(\vect{r},t\right)}\right]}^{\vect{k},\tilde{t}}\right)\nonumber\\
                                                         &+& \gamma\hat{\eta}^{\vect{k},t}, 
\end{eqnarray}
which performs similarly well in many cases, but does not rely on force linearization.

Finally, a third type of predictor-corrector algorithm, ETDPEC, was introduced by Villet and Fredrickson\cite{Villet2010}.
ETDPEC uses exponential time differencing with force linearization for \emph{both} the predictor and corrector steps:
\begin{widetext}
\begin{eqnarray}
  \mu^{\vect{k},\tilde t} & =& \mu^{\vect{k},t} -\left(\frac{1-e^{-\lambda\Delta t\gamma^2\hat{\kappa}\left(\vect{k}\right)}}{\hat{\kappa}\left(\vect{k}\right)}\right)\widehat{\left[\frac{\delta H\left[\left\{\mu\right\}\right]}{\delta \mu\left(\vect{r},t\right)}\right]}^{\vect{k},t} +\left(\frac{1-e^{-\lambda\Delta t\gamma^2\hat{\kappa}\left(\vect{k}\right)}}{2\lambda\Delta t\gamma^2\hat{\kappa}\fnk}\right)\eta^{\vect{k},t}\\
  \mu^{\vect{k},t+\Delta t} &=& \mu^{\vect{k},t} -\frac{1}{2}\left(\frac{1-e^{-\lambda\Delta t\gamma^2\hat{\kappa}\left(\vect{k}\right)}}{\hat{\kappa}\left(\vect{k}\right)}\right)
  \left(\widehat{\left[\frac{\delta H\left[\left\{\mu\right\}\right]}{\delta \mu\left(\vect{r},t\right)}\right]}^{\vect{k},t} + \widehat{\left[\frac{\delta H\left[\left\{\mu\right\}\right]}{\delta \mu\left(\vect{r},t\right)}\right]}^{\vect{k},\tilde t} \right)+\left(\frac{1-e^{-\lambda\Delta t\gamma^2\hat{\kappa}\left(\vect{k}\right)}}{2\lambda\Delta t\gamma^2\hat{\kappa}\fnk}\right)\eta^{\vect{k},t}.
\end{eqnarray}
\end{widetext}

We now discuss the derivation of weak-inhomogeneity linear response kernels, $\hat{\kappa}\left(\vect{k}\right)$, for linear multiblock polymer chains.

\subsection{Weak inhomogeneity expansion}
\label{sec:wie}
A weak inhomogeneity expansion
assumes that local field deviations from a homogeneous saddle-point are weak, 
such that a power-series expansion of the action in a small perturbation parameter can be truncated beyond the quadratic
term. 
The explicit field dependences of the action of both incompressible and weakly 
compressible melt models appear only to second order, but the $\ln Q_p$ term is a 
non-linear, non-local functional containing all powers of the fields.
An expansion of $\ln Q_p$ to second order in the field perturbations yields the following weak inhomogeneity expansion
for the MSE-mapped action of an incompressible melt of \emph{linear} multiblock polymer chains:
\begin{eqnarray}
  H &\approx& H_{0} - \frac{C V}{2}\sum_{\vect{k}\ne 0} \left(\sum_{j=1}^{S-1}\frac{1}{d_j}
  \hat{\mu}_j\left(-\vect{k}\right) \hat{\mu}_j\left(\vect{k}\right)\right.
\nonumber \\
&+& \left.\sum_p \frac{\phi_p}{\alpha_p} 
\sum_{ij=1}^{N_B}\sum_{lm=1}^S \hat{g}^p_{ij}\fnk A_{S_il}\hat{\mu}_{l}\left(-\vect{k}\right)A_{S_jm}\hat{\mu}_{m}\fnk \right),\nonumber\\
\end{eqnarray}
where $H_0$ is evaluated for the homogeneous saddle-point configuration, and we have used periodic boundary conditions with the 
Fourier transform conventions $\hat{f}^\vect{k} = V^{-1}\rint f\fnr \exp\left(-i\vect{k}.\vect{r}\right)$, $f\fnr = \sum_\vect{k} \hat{f}^\vect{k}\exp\left(i\vect{k}.\vect{r}\right)$.
$N_B$ is the number of distinct blocks along the copolymer backbone, and $S_i$ is a length-$N_B$ map vector that specifies the
chemical species index for block $i$.
$\phi_{p}$ is the volume fraction of the $p$th polymer chain type, and $\alpha_p$ is its
chain length relative to $N$.
$g^p_{ij}({\bf k})$ is a Debye pair correlation factor in
reciprocal space applied between blocks $i$ and $j$ of uniform species index on polymer chain type $p$:
\begin{equation}
\hat{g}^p_{ij}\fnk = \left\lbrace{\begin{array}{lr}
    \hat{g}\left(\left|\vect{k}\right|,f_i^p\right) , & i = j\\
    \hat{h}\left(\left|\vect{k}\right|,f_i^p\right) \hat{h}\left(\left|\vect{k}\right|,f_j^p\right) \hat{l}\left(\left|\vect{k}\right|,d_{ij}^p\right), & i \ne j
\end{array}} \right.
\end{equation}
with the scalar-valued functions
\begin{eqnarray}
  \hat{g}\left(k,f\right) &=& 2k^{-4} \left( e^{-fk^2} + fk^2 - 1 \right)\\
  \hat{h}\left(k,f\right) &=& k^{-2}\left(1 - e^{-fk^2}\right) \\
  \hat{l}\left(k,d\right) &=& e^{-d k^2}
\end{eqnarray}
where $f^p_{i}$ is the normalized contour length (such that $\sum_i f^p_i = 1\, \forall\, p$ regardless of $\alpha_p$) of
a sequential block of segments of species $i$ on chain $p$, and $d_{ij}$ is the sum of all normalized block lengths 
between, but not including, blocks $i$ and $j$ along the chain contour.
Notice that $\hat{g}_{ij}^p\left({\bf k}\right)$ is only evaluated
for block pairs $(i, j)$ belonging to the same polymer component $p$.
Higher-order inter-chain terms do not appear, because those are handled by
the explicit interaction terms in the action, while $\ln Q_p$ is the normalized partition
function of non-interacting chains in external fields.
Similar expressions can be derived for branched polymer chain architectures using
perturbation theory\cite{fredrickson06}.

For a weakly compressible melt, the weak inhomogeneity expansion is:
\begin{eqnarray}
  H &\approx& H_{0} - \frac{C V}{2}\sum_{\vect{k}\ne 0} \left(\sum_{j=1}^{S}\frac{\left(\zeta N\right)^{-\frac{1}{2}}}{d_j}
  \hat{\mu}_j\left(-\vect{k}\right) \hat{\mu}_j\left(\vect{k}\right)\right.
\nonumber \\
&+& \left.\sum_p \frac{\phi_p}{\alpha_p} 
\sum_{ij=1}^{N_B}\sum_{lm=1}^S \hat{g}^p_{ij}\fnk O_{S_il}\hat{\mu}_{l}\left(-\vect{k}\right)O_{S_jm}\hat{\mu}_{m}\fnk \right),\nonumber\\
\label{eqn:wie_comp}
\end{eqnarray}
with the same Debye correlation functions as in the incompressible model.

\subsection{Linearized derivatives}
The linearized forces employed in Sec.\ \ref{sec:numerics}A can be obtained by functional derivatives of the 
second-order weak-inhomogeneity expansions provided in the previous section. For the incompressible model:
\begin{eqnarray}
  \widehat{\left[\frac{\delta H\left[\left\{\mu\right\}\right]}{\delta \mu_{l}\fnr}\right]}^\vect{k}_{lin} &=&
  -\frac{C\hat{\mu}_l\fnk }{d_{l}} \nonumber \\
  & -& \sum_p \frac{C\phi_p}{\alpha_p} \sum_{ij=1}^{N_B}\sum_{m=1}^{S} \hat{g}^p_{ij}\fnk A_{S_il} A_{S_jm} \hat{\mu}_m\fnk \nonumber \\
  \widehat{\left[\frac{\delta H\left[\left\{\mu\right\}\right]}{\delta \mu_{+}\fnr}\right]}^\vect{k}_{lin} &=& - \sum_p \frac{C\phi_p}{\alpha_p} \sum_{ij}^{N_B}\sum_{m=1}^{S} \hat{g}^p_{ij}\fnk A_{S_i+} A_{S_jm} \hat{\mu}_m\fnk. \nonumber \\
\label{linderiv}
\end{eqnarray}

For the purposes of stabilizing the time stepping algorithms presented earlier, 
it is advantageous to only consider those parts of 
the linearized derivatives that are proportional to the field whose derivative
is calculated; contributions proportional to other fields are in general not
stabilizing. Hence,
\begin{equation}
  \widehat{\left[\frac{\delta H\left[\left\{\mu\right\}\right]}{\delta \mu_{i}\fnr}\right]}^\vect{k}_{lin} \approx \hat{\kappa}_{i}\fnk \mu_{i}\fnk, 
\label{linderiv_num}
\end{equation}
where $\hat{\kappa}$ is a linear-response kernel.
Moreover, instead of the full functions $\hat{\kappa}_{i}\fnk$ we retain only those
terms whose sign suggest that they are stabilizing for implicit time stepping, resulting in
\begin{eqnarray}
  \label{eqn:kappaj}
  \hat{\kappa}_{j}\fnk &=& -\frac{C}{d_{j}}\theta\left(-d_i\right) -\sum_{p}\frac{C\phi_p}{\alpha_p}\sum_{il=1}^S \hat{g}^p_{il}\fnk A_{ij}A_{lm}\theta\left(A_{ij}A_{lm}d_{j}\right)\nonumber\\\\
  \hat{\kappa}_{+}\fnk &=& -\sum_p \frac{C\phi_p}{\alpha_p}\hat{g}\left(\left|\vect{k}\right|,\alpha_p\right)
  \label{eqn:kappa+}
\end{eqnarray}
with the Heaviside function $\theta\left(x\right)=1$ for $x>0$ and $=0$ otherwise.
The $\theta$ terms are introduced so that only terms with the correct sign for stable implicit time stepping are included in $\kappa$.
For the results presented in Section \ref{sec:results}, we use only $\hat{\kappa}_+$, together with all explicitly linear terms, 
for implicit time stepping.
Further improvements in time stepping stability and accuracy might be achieved by also using Eqn.\ \ref{eqn:kappaj}.
Analogous expressions can be developed for the compressible model, though the model in general lacks a pure pressure-like
field analogous to $\mu_+$.
Our present strategy mirrors the non-pressure modes of the compressible model: we take only the stabilizing explicitly linear 
terms for implicit time stepping.

\section{Results}
\label{sec:results}
\subsection{Mean-field Theory}
As discussed in Section \ref{sec:theory}A, the complex Langevin dynamics 
used for sampling the field theoretic partition function will return
mean-field saddle-point solutions if the driving noise is eliminated.
This approximation becomes asymptotically exact for $C\rightarrow\infty$, a limit for
which the noise strength is naturally vanishing\cite{ghf87,ganesan01,lennon08} 
relative to $\delta H/\delta \mu$.
Although the primary focus of the present work is the development of stable schemes
for propagating CL dynamics, as a first test of the correctness of the 
MSE mapping we here explore mean-field simulations for linear ABC triblock 
melts and compare to known results.

\paragraph*{Consistency of $ABC$ mean-field phase diagram: }
Throughout this work we consider only conformationally symmetric systems (that is, those for which 
the statistical segment lengths are all equal to the reference $b$); this restriction is easily lifted.
The parameter space for incompressible melts of linear, monodisperse triblock
polymers is then five dimensional: two block fractions ($f_A$, $f_B$, and $f_C=1-f_A-f_B$), 
and three $\chi N$ parameters ($\chi_{AB}N$, $\chi_{AC}N$, $\chi_{BC}N$). 
For a three-species incompressible melt, the reduced interaction matrix 
$X_{ij} = \chi_{ij}N-2\chi_{iS}N$ becomes a $2\times 2$ matrix. 
If we further restrict our initial tests to the case 
of $\chi N = \chi_{AB}N = \chi_{AC}N = \chi_{BC}N$, we find
\begin{eqnarray}
  \mat{X} &= &\chi N\left(
    \begin{array}{c c}
      -2 & -1 \\
      -1 & -2 \\
    \end{array}
  \right),\\
  \mat{O} & = &\frac{1}{\sqrt{2}}\left(
    \begin{array}{c c}
      -1 & -1 \\
      -1 & 1  \\
    \end{array}
  \right),\\
  d_1 & = & -3\chi N, \quad d_2 = -\chi N.
\end{eqnarray}
These quantities fully describe the MSE mapping and the meaning of the 
exchange-mapped fields $\mu_1\fnr, \mu_2\fnr, \mu_+\fnr$ for this specific system.

\begin{figure}[H!t]
\begin{center}
{\includegraphics[width=0.9\columnwidth]{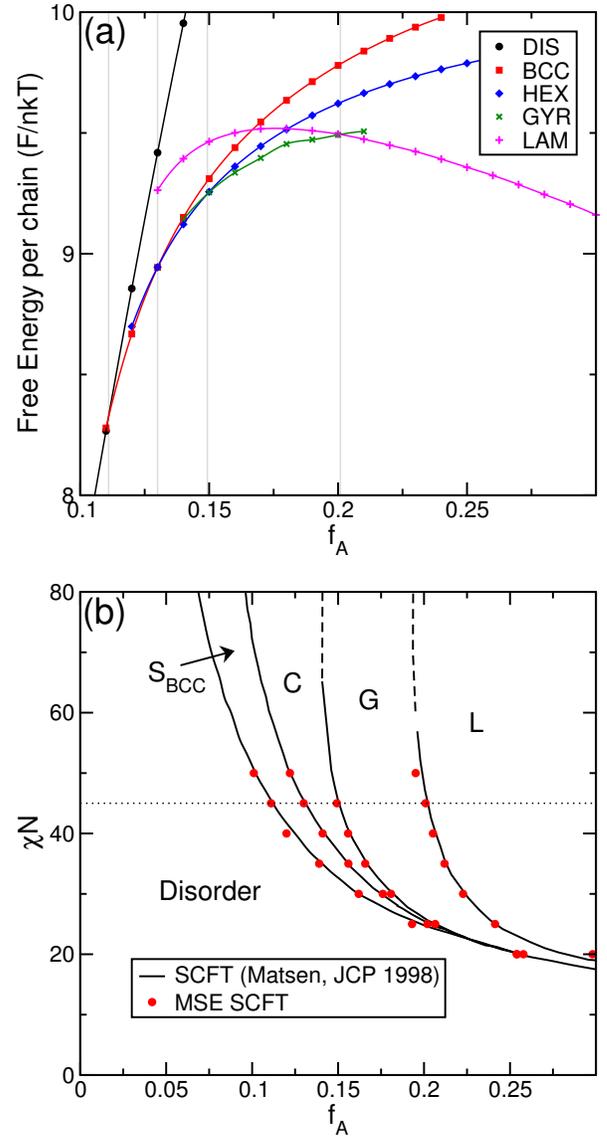}}
\caption{Phase diagram for an incompressible symmetric ABC triblock melt
with $\chi N = \chi_{AB}N = \chi_{BC}N = \chi_{AC}N$.
Panel (a) shows the SCFT free energy, computed as $H$ of Eqn.\ \ref{eqn:H_incomp_mse_body} at 
the saddle-point field configuration ($\left\{\mu^\star\right\}$) for each phase under consideration, plotted against the A block fraction $f_A (=f_C)$ at $\chi N = 45$. 
Lines are cubic-spline interpolants.
Phase transitions are shown as vertical grey lines. Panel (b) shows a reproduction of Matsen's phase diagram\cite{matsen98b}
with selected phase boundary points from the MSE-SCFT method denoted as red circles. The dashed line shows the cut at $\chi N = 45$ 
used to generate panel (a).
}\label{fig:SCFT_ABCPhases}
\end{center}
\end{figure}

A partial SCFT phase diagram was first presented by Matsen\cite{matsen98b} for symmetric triblock polymers ($f_A = f_C$).
As a first test of the correctness of the MSE mapping, Matsen's ABC phase diagram is partially reproduced in Fig.\ \ref{fig:SCFT_ABCPhases}. 
Excellent agreement between the MSE-SCFT results and Matsen's species-model SCFT phase diagram is observed.

\begin{figure}[H!t]
\begin{center}
{\includegraphics[width=0.9\columnwidth]{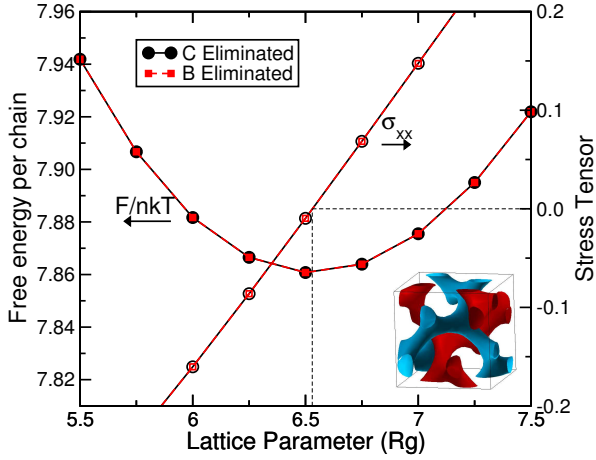}}
\caption{Helmholtz free energy per chain and stress tensor as a function of cubic lattice parameter for the
alternating double-gyroid phase with space group $I4_132$, demonstrating the invariance
of thermodynamic observables to the species that is eliminated in the incompressible MSE mapping. 
Calculations were conducted at $\chi N = 35$ and $f_A = f_C = 0.19$.
The stress tensor is related to the derivative of the intensive free energy with respect to lattice parameters at
\emph{constant concentration}, and is used as the relevant operator in our variable-cell method\cite{barrat05}.
The stress-free configuration has lattice parameter $6.53$\,$R_g$. Inset shows isosurfaces of $A$ and $C$ normalized 
density equal to $0.85$.}
\label{fig:SCFT_HvsV}
\end{center}
\end{figure}

\paragraph*{Invariance of thermodynamic properties to eliminated species: }
A feature of the MSE model of an incompressible melt is the elimination of $\hat{\rho}_S$ by employing the
delta functional subspace projection.
An important test case then is to demonstrate that the thermodynamic properties computed from SCFT and CL simulations 
are invariant to the choice of species that is eliminated. 
For example, in an ABC triblock melt, the C ($=S$) species, being the highest indexed, is eliminated by default, but we are free
to eliminate any other by permuting the species labels.
This test, shown in Fig.\ \ref{fig:SCFT_HvsV}, is made with SCFT calculations of the alternating gyroid phase of an ABC triblock melt, using the free-energy variation with 
lattice parameter and the stress tensor as example thermodynamic properties.

\begin{figure}[H!t]
\begin{center}
{\includegraphics[width=0.8\columnwidth]{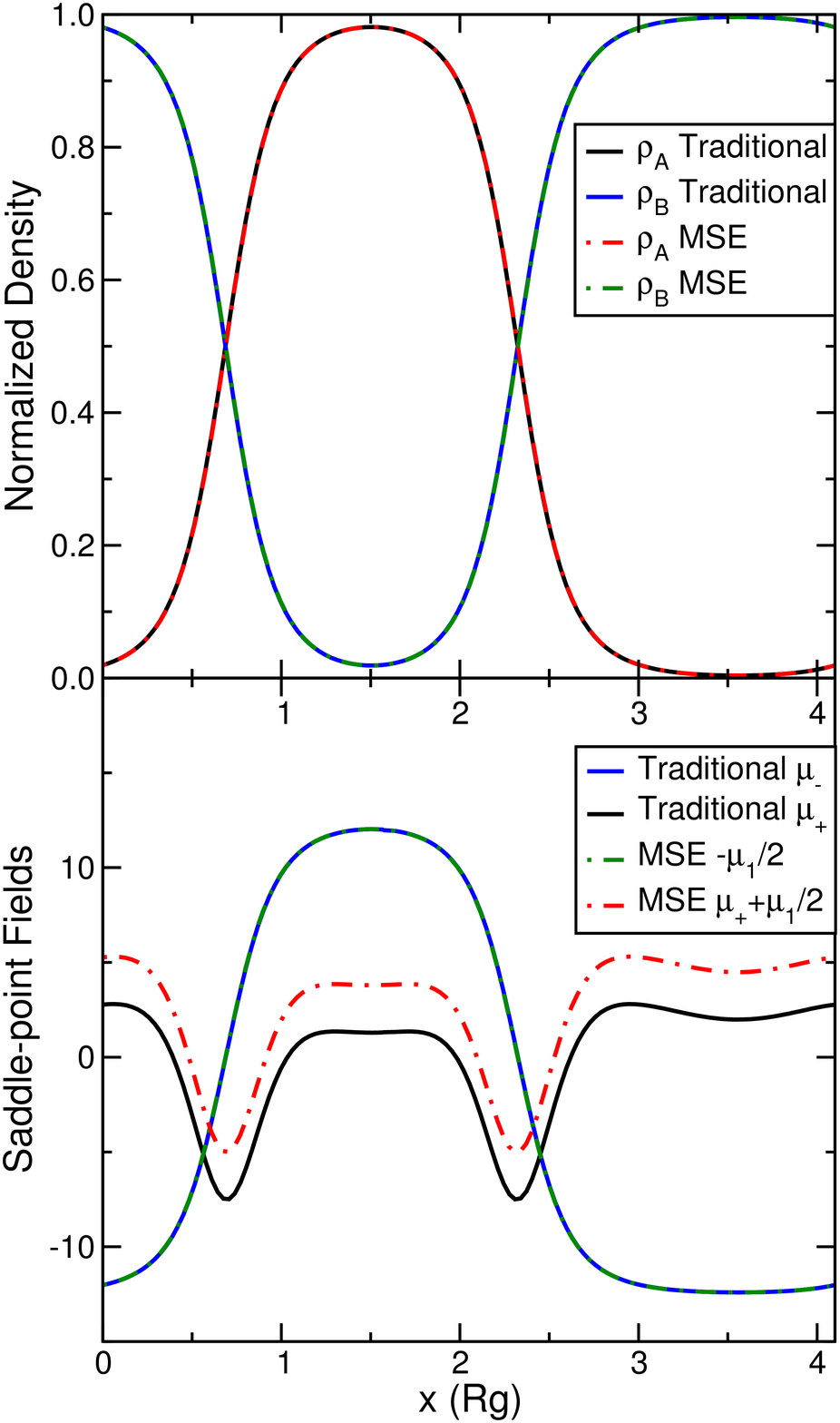}}
\caption{Density profiles (upper panel) and exchange field configurations (lower panel) for
  mean-field configurations of an incompressible AB diblock copolymer melt in the lamellar phase ($\chi N=25$, $f_A=0.4$) using both
the traditional\cite{fredrickson06} two-species exchange mapping (dashed lines) and MSE (solid lines).
The density profiles match between the two methods. The fields are in agreement with
the appropriate mapping (see text) apart from an arbitrary shift of the pressure field, as
discussed in Section \ref{sec:theory}A, CL.}
\label{fig:2specreduction}
\end{center}
\end{figure}

\paragraph*{Reduction to two-species exchange-mapped field theory: }
As a further test of the MSE mapping, we consider the special case of a system consisting of two species.
In this case the interaction matrix $X_{ij}$ has only one entry $X_{11}=-2\chi_{AB}N$.
The parameters that enter the action are therefore $d_1 = -2\chi_{AB}N$ and $O_{11} = 1$, and the action itself is 
\begin{eqnarray}
  H &=& C\left[\frac{1}{4\chi_{AB}N}\rint\mu_1\fnr - \rint\left(\mu_+\fnr + \frac{\mu_1\fnr}{2}\right)\right.\nonumber\\
&-&\left.\sum_{p=1}^PV\frac{\phi_p}{\alpha_p}\ln Q_p\right].
\end{eqnarray}
The species chemical potential fields that enter the $Q_p$ functionals 
are defined in terms of the exchange fields through the $A_{ij}$ matrix
as $\psi_A\fnr N = \mu_1\fnr+\mu_+\fnr$, and $\psi_B\fnr N = \mu_+\fnr$.
The traditional two-species exchange-mapped field theory\cite{fredrickson06}
has the relationships $\psi_A\fnr N = \mu_+^{(2)}\fnr+\mu_-^{(2)}\fnr$, and $\psi_B\fnr N = \mu_+^{(2)}\fnr-\mu_-^{(2)}\fnr$.
We therefore expect to find consistent saddle-point field configurations between 
the two methods if we identify $\mu_+\fnr + \mu_1\fnr/2 = \mu_+^{(2)}$ and 
$-\mu_1\fnr/2 = \mu_-^{(2)}$.
Fig.\ \ref{fig:2specreduction} shows such an example for the saddle-point solutions of
an AB diblock copolymer melt in the lamellar phase.
The normalized species densities (upper panel) are immediately equal between the reduced MSE and 
traditional two-species exchange mapping (as are scalar thermodynamic quantities such as the free energy
and components of the stress tensor), but the auxiliary exchange chemical potential fields (lower panel) require the mapping above to be brought  
into agreement.
This test verifies that MSE correctly reduces to the traditional two-species exchange mapping in the appropriate limit, up to an 
arbitrary shift in the $\mu_+\fnr$ field.

\begin{figure}[H!t]
\begin{center}
{\includegraphics[width=0.9\columnwidth]{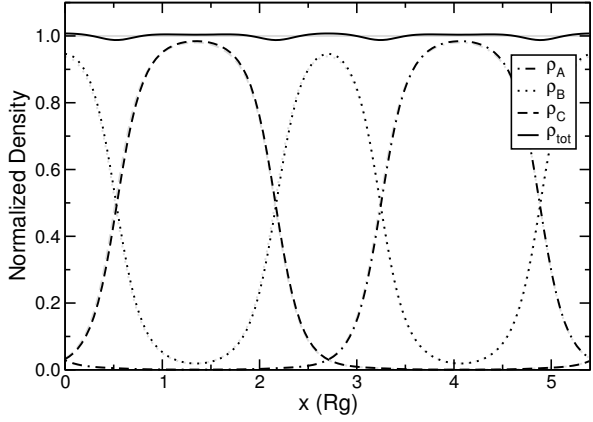}}
\caption{Cross-sectional profiles of SCFT monomer densities for an incompressible melt (grey) and weakly compressible melt (black), with Helfand penalty $\zeta N=100$,
of an ABC triblock polymer melt in the lamellar phase. The system parameters are $\chi_{AB}N = \chi_{BC}N = \chi_{AC}N = 30$ and $f_A = f_C =0.3$.}
\label{fig:scftcompressible}
\end{center}
\end{figure}

\begin{figure}[H!t]
\begin{center}
{\includegraphics[width=0.9\columnwidth]{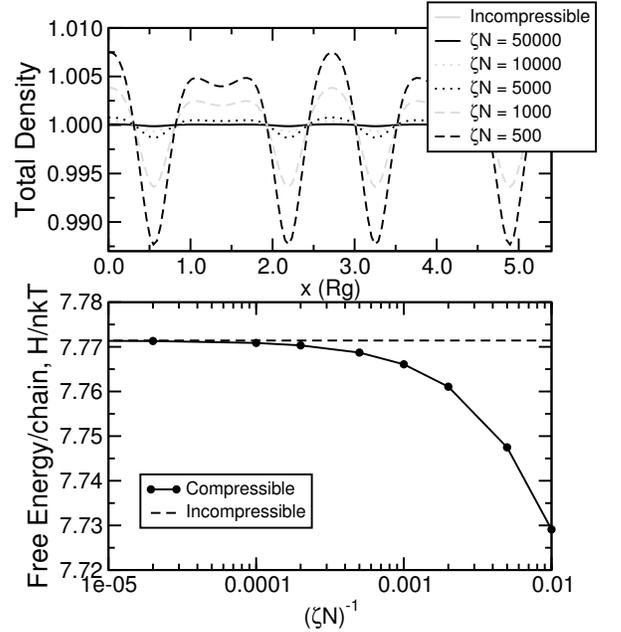}}
\caption{The same system as in Fig.\ \ref{fig:scftcompressible}. Upper panel: sections of total density profiles ($\rho_A\fnr + \rho_B\fnr + \rho_C\fnr$)
approach the incompressible limit ($=\rho_0$ for all $\vect{r}$) as the $\zeta N$ penalty is increased. Lower panel: the SCFT free energy per chain continuously approaches
that generated by the fully incompressible model as $\zeta N$ is increased.}
\label{fig:scftcompressible_incomplimit}
\end{center}
\end{figure}

\paragraph*{Finite compressibility: }
Here we demonstrate the effect on SCFT morphologies of permitting weak violation of the incompressibility condition by the introduction of 
a Helfand compressibility penalty.
Figure \ref{fig:scftcompressible} demonstrates the monomer density profiles in the saddle-point approximation of an ABC triblock polymer melt in the
lamellar phase.
The compressible model (black lines) has a non-uniform total density $\hat{\rho}_A\fnr + \hat{\rho}_B\fnr + \hat{\rho}_C\fnr \neq \rho_0$, 
with strongest violations in the vicinity of interfaces. 
Fig. \ref{fig:scftcompressible_incomplimit} shows the continuous approach to the incompressible limit of both the free energy and total
density profiles of the same ABC triblock melt.
Relaxing the incompressibility constraint can be numerically advantageous in a number of situations, especially in cases for which the 
incompressible constraint is difficult to satisfy, such as in systems \emph{constrained} to have non-uniform total density, e.g., due to 
boundary conditions.

\paragraph*{Efficiency of the saddle-point search in a compressible model:}
\label{sec:scft_compressible_results}
For simplicity we make our analysis of time stepping efficiency 
for the same system as in Fig.\ \ref{fig:scftcompressible}: an 
ABC linear triblock polymer melt in the lamellar phase 
with all binary $\chi N$ interactions equal.
This specific choice makes the interaction matrix, $\mat{X} = \left(\mat{\chi N}\left(\zeta N\right)^{-\frac{1}{2}} + \left(\zeta N\right)^{\frac{1}{2}}\mat{\openone}\right)$
introduced in Appendix \ref{sec:appendix_comp}, have one eigenmode equal to $\left(1,\ldots,1\right)$, mirroring the character of the pure pressure field in the incompressible
model. 
Furthermore, the $\mat{X}$ matrix has only two distinct eigenvalues: $d_i = -\chi N/\sqrt{\zeta N}$ for $i \in \left[1,S-1\right]$, and 
$d_S = \left(S\zeta N + \left(S-1\right)\chi N\right)/\sqrt{\zeta N}$ for the pressure-like mode.
Consequently, $d_S$ diverges as the incompressible limit is approached, while all other $d_i \rightarrow 0$.
This expansion of the eigenvalue spectrum leads to a critical slowing of the relaxation rate of the compressible model as the incompressible limit is approached.

\begin{figure*}[H!t]
\begin{center}
{\includegraphics[width=0.25\textwidth]{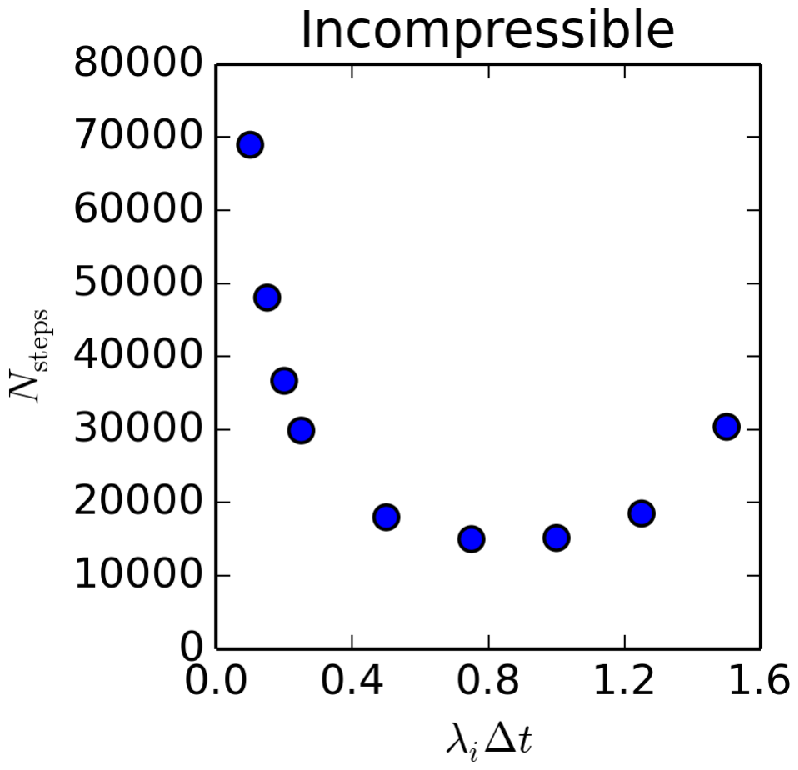}
\includegraphics[width=0.74\textwidth]{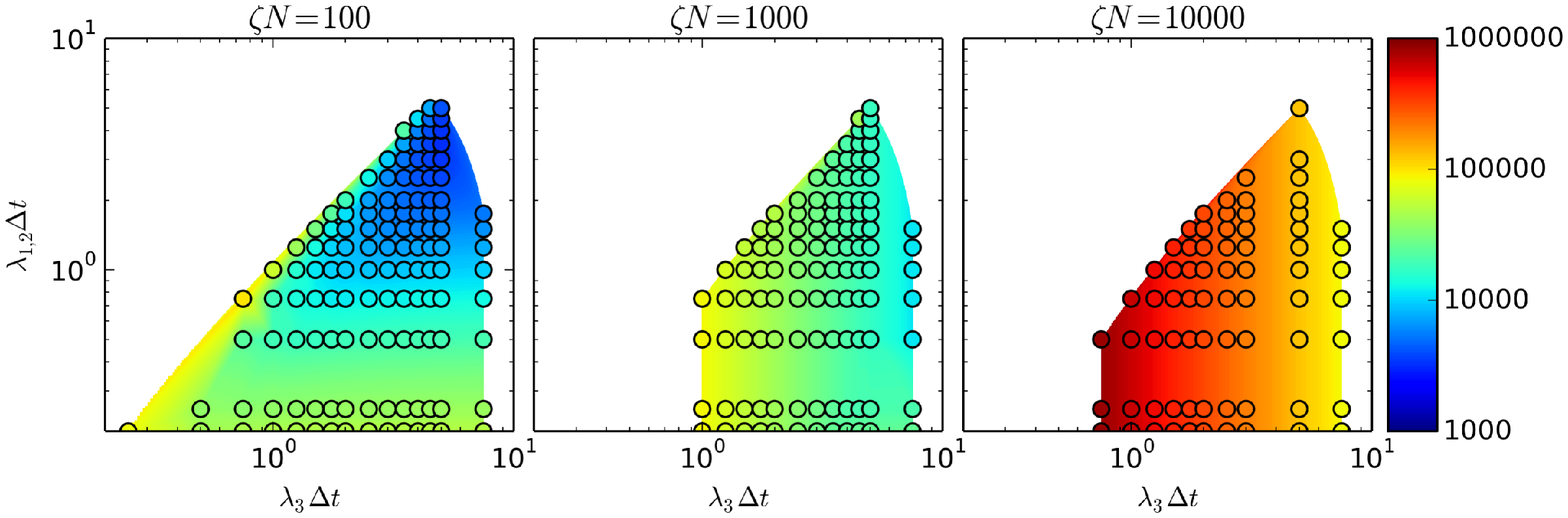}}
\caption{(Color online) Number of time steps required to reach the saddle-point solution as scalar mobility coefficients are
  varied for an incompressible system (left) and three different compressibile systems. The system is an ABC linear-chain triblock melt in the lamellar
phase with all $\chi N=30$ and $f_A = f_C = 0.3$. The SIS field updater was used for all calculations. 
For the incompressible model, the mobilities of all modes are varied together, mirroring typical 
use. For compressible models, the mobilities were partitioned into variation for the 
pure pressure-like mode, $\lambda_3$, and for the remaining modes, $\lambda_1=\lambda_2$, mirroring the separation
of MSE-mapping eigenvalues. Circles show convergent simulations, with background shading obtained via interpolation. The unshaded regions
had no convergent simulations.}
\label{fig:comp_convergence_map}
\end{center}
\end{figure*}

Studying the form of Equations \ref{eqn:comp_eom} and \ref{eqn:comp_forces} indicates that scaling the mobility $\lambda_i \approx d_i$ might largely compensate for
the spectrum broadening.
However, this proposed startegy is oversimplified, because the density fields $\varphi_i\fnr$, or equivalently the single-chain conformational term $\ln Q$, entangles 
all modes to all orders, as revealed even in the lowest-order expansion of the forces in Eqn.\ \ref{eqn:wie_comp}.
Lacking a general strategy for efficient selection of mobility coefficients, we instead derive understanding by studying rates of convergence numerically.
Figure \ref{fig:comp_convergence_map} shows an analysis of a map of the number of mean-field relaxation steps required to converge to the lamellar saddle-point solution
as a function of $\lambda_i$ mobility coefficients.
Mirroring the separation of the eigenvalue spectrum, we explore scalar-valued mobility coefficients $\lambda_i \ne \lambda_S$ with $\Delta t=1$, and
determine the stability envelope as well as the number of SCFT iterations required to reach the saddle point configuration.

For the incompressible model, the fastest convergence to the saddle-point configuration (to a tolerance of 
less than $10^{-6}$ in the L2 norm of all forces)
was achieved with $\lambda_i \Delta t=0.75$, with approximately $15,000$ iterations of the SIS field updater.
For moderately larger $\lambda_i \Delta t$, the number of iterations is increased due to a damped oscillatory approach to the solution. 
Further increases in the mobility coefficient lead to non-damped oscillations or divergent trajectories.
From the compressible data, the following observations can be made:
\begin{itemize}
  \item Almost without exception, the pressure-like mode, $\mu_3$, required a mobility at least as large as the other modes for stable field trajectories.
  \item For weak compressibility penalty (more compact eigenvalue spectrum), optimal performance can be achieved with $\lambda_3 \approx \lambda_{1,2}$. In contrast,
    strong compressibility penalties (wide eigenvalue spectrum) favor a strong asymmetry in mobility coefficients.
    In that case, increasing $\lambda_i$ for $i\in\left[1,S-1\right]$ does not significant increase the rate of convergence; increasing $\lambda_S$ does.
  \item With an appropriate choice of $\left\{\lambda_i\right\}$, the inefficiency related to the sparse eigenvalue spectrum on the approach to
    the incompressible limit is significantly ameliorated, while a poor choice leads to either increased computational cost or
    instability.
  \item If $\lambda_i\Delta t=1.00$ is chosen, the $\zeta N=100$, $1,000$ and $10,000$ models required respectively $\sim 62,000$, $\sim 70,000$, and $\sim 510,000$ iterations of the SIS field updater.
    In contrast, tuning to the optimal mobility coefficients in the space of $\lambda_1 = \lambda_2 \neq \lambda_3$ results in a minimum iteration count of
    $\sim 2,500$ for $\zeta N=100$, $\sim 11,500$ for $\zeta N=1,000$, and $\sim 85,000$ for $\zeta N=10,000$, at least a factor $5$ improvement.
  \item Fig.~\ref{fig:scftcompressible_incomplimit} shows that $\zeta N=1000$ differs from the incompressible limit by $\sim 0.05\%$ in the SCFT free energy and by a peak of
    $\sim 1\%$ in the spatial variation of the total segment density. The $\zeta N=1,000$ model required approximately 30\% fewer time steps than the incompressible model in this test case.
\end{itemize}

\subsection{Complex Langevin Simulations}
\begin{figure*}[H!t]
\begin{center}
{\includegraphics[width=\textwidth]{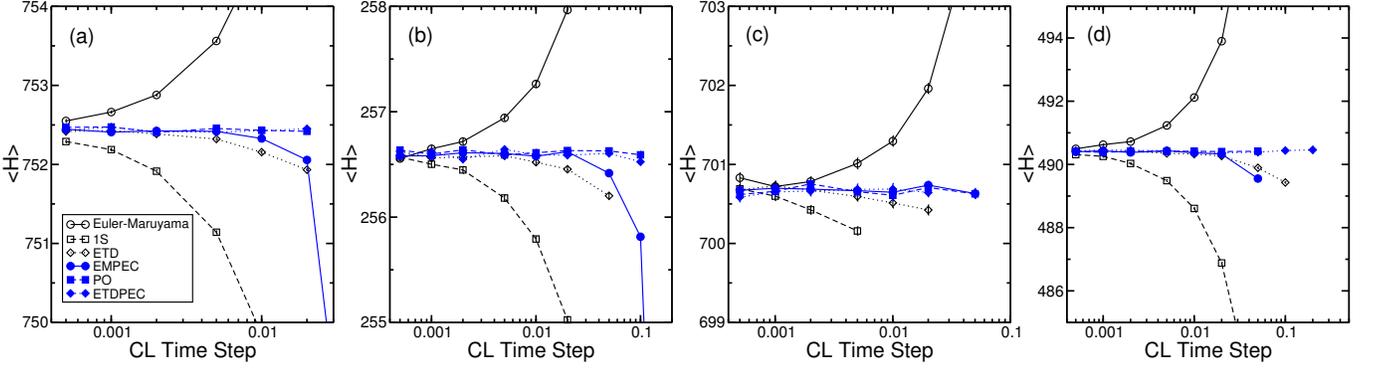}}
\caption{Influence of time step size on the value of $\left<H\right>$ obtained from CL simulations.
  Different time integration algorithms are compared (see text).
(a) A symmetric ABC triblock melt in the disordered phase with $f_B=0.6$, $\chi_{AB} N = \chi_{BC} N = 10$, $\chi_{AC}N=14$ and $C=100$. (b) As panel (a), 
with $C=30$ (stronger fluctuations). (c) The same melt at $\chi_{AB}N = \chi_{BC}N=\chi_{AC}N = 30$ and $C=50$ in the alternating gyroid phase (Fig.\ \ref{fig:SCFT_HvsV}). (d)
A symmetric ABDC tetrablock melt in the disordered phase with $f_A = f_C = 0.2$, $f_B = f_D = 0.3$. Neighboring blocks on the chain interact with $\chi N = 10$, non-neighbors with $\chi N=14$.
Filled symbols are predictor-corrector algorithms, and open symbols are single-shot time integrators.
Standard errors of the mean estimate, plotted as vertical error bars, are significantly smaller than the symbol size in all panels except (c). 
In order to keep the error-bar size consistent with reducing $\Delta t$, we scale the total number of CL steps (and hence the total simulation time) inversely.
}
\label{fig:CLDT}
\end{center}
\end{figure*}

\paragraph*{Time-integration algorithms: }
Figure \ref{fig:CLDT} shows the variation of the thermal average of a trial operator, $\left<H\right>$, as the time integration
algorithm and time-step size are both varied for a selection of multi species incompressible melts.
Only simulations that did not yield divergent trajectories have data included.
A crucial finding is that all methods are stable (no divergent trajectories observed) and 
accurate (time-step error controlled compared to statistical uncertainties) for sufficiently small values of $\Delta t$.
However, working with small $\Delta t$ is very inefficient, because the correlation time, and therefore the number of discrete time steps required to move 
between statistically distinct configurations, is long. 
The number of statistically decorrelated configurations used in computing thermal averages is a 
crucial quantity for finding the standard error of the mean, and therefore determining stochastic
uncertainty on predictons of observables. 
The standard error of the mean decreases as the square root of the number of decorrelated field configurations sampled.
For Fig.\ \ref{fig:CLDT}, simulations of the incompressible model were ran long enough to yield standard errors smaller than or equal to the
symbol size (i.e., smaller $\Delta t$ simulations completed more time steps).
Stability and accuracy trends are similar for the compressible model.
%

Fig.\ \ref{fig:CLDT} shows that Euler-Maruyama and 1S methods, while quite stable at the $\Delta t$ range of interest, are least effective from the viewpoint of sampling efficiency due to a large time-step
bias.
Exponential time differencing shows significant improvements in accuracy, permitting approximately one order of magnitude improvement in efficiency.
Each of these methods has essentially equal computational cost per CL time step, which is dominated by solving the modified diffusion equations for chain propagators.
EMPEC, the simplest predictor-corrector method, improves accuracy further while doubling the runtime per CL time step.
Finally, Petersen-\"{O}ttinger and ETDPEC methods demonstrate univerally excellent accuracy in all tests, with 
time step sizes being limited by stability rather than accuracy.
Table \ref{tab:FUData} summarizes a quantitative comparison of sampling efficiency for each time-step method.
It is evident that predictor-corrector methods are crucial for efficient access to advanced applications of CL sampling.

\begin{table*}[htbp]
  \centering
  \begin{tabular}{c|c||crcrr}
    & Time Stepper & $\Delta t_\mathrm{max}$\footnote{$\Delta t$ required to converge $\left<H\right>$ to within 0.05\% of low-$\Delta t$ value.} & $N_t^\mathrm{min}$\footnote{$N_t\left(\Delta t_\mathrm{max}\right)$ is the number of time steps required to reduce the standard error of the mean ($\epsilon$) sufficiently so that $3\epsilon$ is no larger than $1$ in the 4th significant figure of $\left<H\right>$} & $\left<H\right>$ & $\tau_\mathrm{corr}\left(\Delta t_\mathrm{max}\right)$ & runtime (sec) \\
    \hline
    \hline
    & EM & 0.002 & 16500 & 265.72(3) & 420 & 760\\
    \cline{2-2}
    Linear ABC  & 1S & 0.002 & 20000 & 256.45(3) & 630 & 880\\
    \cline{2-2}
    DIS phase  & ETD & 0.020 & 2450 & 256.45(3) & 85 & 108\\
    \cline{2-2}
    $\chi N=10,14,10$ & EMPEC & 0.050 & 1250 & 256.41(3) & 70 & 103\\
    \cline{2-2}
  $C=30$ & PO & 0.100\footnotemark[3] & 500 & 256.59(3) & 60 & 44\\
    \cline{2-2}
    & ETDPEC & 0.100\footnotemark[3] & 500 & 256.52(3) & 55 & 44\\
    \hline
    & EM & 0.005 & 180000& 700.98(3) & 260 & 12200 \\
    \cline{2-2}
    Linear ABC  & 1S & 0.002 & 321500 & 700.41(3) & 400 & 16600 \\
    \cline{2-2}
    GYR phase  & ETD & 0.020 & 49000 & 700.42(3) & 65 & 3580 \\
    \cline{2-2}
    $\chi N=30,30,30$ & EMPEC & 0.050\footnotemark[3] & 18000 & 700.61(3)& 60 & 1465\\
    \cline{2-2}
    $C=50$ & PO & 0.050\footnotemark[3] & 13600 & 700.63(3) & 50 & 1320 \\
    \cline{2-2}
    & ETDPEC & 0.020\footnotemark[3] & 33850 & 700.64(3)& 65 & 3260\\
    \hline
  \end{tabular}
    \footnotetext[3]{Unstable before inaccurate}
  \caption{Quantitative comparison of CL time stepper performance during a single calculation.
    Since CL is a stochastic method, some variability in the data should be expected in repeated trials.
    The time step is chosen to be equal to $\Delta t_\mathrm{max}$, defined as the time step required to converge
    the estimate of $\left<H\right>$ to within 0.05\% of the low-$\Delta t$ value. For the chosen time step, each simulation ran
    for a number of time steps required to make the error of the mean sufficiently small to resolve four significant figures.
    Test systems were a disordered phase at low $C$ with asymmetric $\chi N$ parameters, and an alternating gyroid phase
    at moderate $C$ with symmetric $\chi N$ parameters. Both phases employed a spatial collocation mesh of $32^3$, second-order
    operator splitting for solving the modified diffusion equation with contour discretization $\Delta s=0.01$, and a periodic
    simulation cell with a volume of $512 R_g^3$ for the disordered phase and $303.5 R_g^3$ for the gyroid phase.
    Runtimes are for a single NVIDIA Tesla C2070 GPU\cite{Delaney/Fredrickson:2013}.
    Scalar operators are block-averaged over blocks of 50 time steps, so $N_t$ is required to be a multiple of 50 in these tests.  }
  \label{tab:FUData}
\end{table*}

Now that we have demonstrated controlled stability and accuracy of the stochastic sampling, we address the physical correctness of the CL-MSE method. 

\paragraph*{Structure factors: }
One physical test of the CL-MSE simulations involves comparing the melt structure factors in the disordered phase 
to the RPA predictions of the same quantity. 
Since RPA structure factors contain fluctuations to Gaussian order, the numerical and analytical predictions should be in good agreement, provided the
fluctuations are not sufficiently strong to engage higher-order terms.
This will be the case for high molecular-weight melts with weak interactions far from an order-disorder transition.

The structure factor components are defined for translationally invariant systems as
\begin{eqnarray}
  S_{ij}\fnk = \frac{1}{V}\int d\vect{r}\int d\vect{r}^\prime e^{-i\vect{k}.\left(\vect{r}-\vect{r}^\prime\right)}\left<\delta\hat{\rho}_i\fnr\delta\hat{\rho}_j\left(\vect{r}^\prime\right)\right>\quad
\end{eqnarray}
where $\delta \hat{\rho}_i\fnr = \hat{\rho}_i\fnr - \left<\hat{\rho}_i\fnr\right>$, and $i, j\in\left[1,S\right]$ are species labels.
Using our Fourier transform conventions detailed in Section \ref{sec:theory}B, the structure factors become
\begin{eqnarray}
  S_{ij}\fnk = V\left<\delta\hat{\rho}_i\left(\vect{k}\right)\delta\hat{\rho}_j\left(-\vect{k}\right)\right>.
\end{eqnarray}
Since the microscopic density operators $\hat{\rho}_i$ are integrated out of our field theory, we must use the
methods described in Reference \onlinecite{fredrickson06} to derive an appropriate field-based structure-factor operator.
%
We arrive at
\begin{eqnarray}
  \frac{S_{ij}\fnk}{\rho_0 N} & = & CV\left<\sum_{k=1}^{S-1} \frac{A^{-1}_{ki}\hat{\mu}_k\fnk}{d_k}\hat{\varphi}_j\left(-\vect{k}\right)\right>\nonumber\\
                              & - & CV\left<\sum_{k=1}^{S-1} \frac{A^{-1}_{ki}\hat{\mu}_k\fnk}{d_k}\right>\Bigg<\hat{\varphi}_j\left(-\vect{k}\right)\Bigg>
\end{eqnarray}
for incompressible systems, and
\begin{eqnarray}
  \frac{S_{ij}\fnk}{\rho_0 N} & = & CV\left<\sum_{k=1}^{S} \frac{O_{ik}\hat{\mu}_k\fnk}{\left(\zeta N\right)^{1/2}d_k}\hat{\varphi}_j\left(-\vect{k}\right)\right>\nonumber\\
                              & - & CV\left<\sum_{k=1}^{S} \frac{O_{ik}\hat{\mu}_k\fnk}{\left(\zeta N\right)^{1/2}d_k}\right>\Bigg<\hat{\varphi}_j\left(-\vect{k}\right)\Bigg>
\end{eqnarray}
for the compressible case, where $\varphi_j\fnr$ is defined in Eqn.\ \ref{rho_i}.
These expressions were derived using a single functional integration by parts, which yields the least noisy estimator out of those tested.
%

\begin{figure}[H!t]
\begin{center}
{\includegraphics[width=0.9\columnwidth]{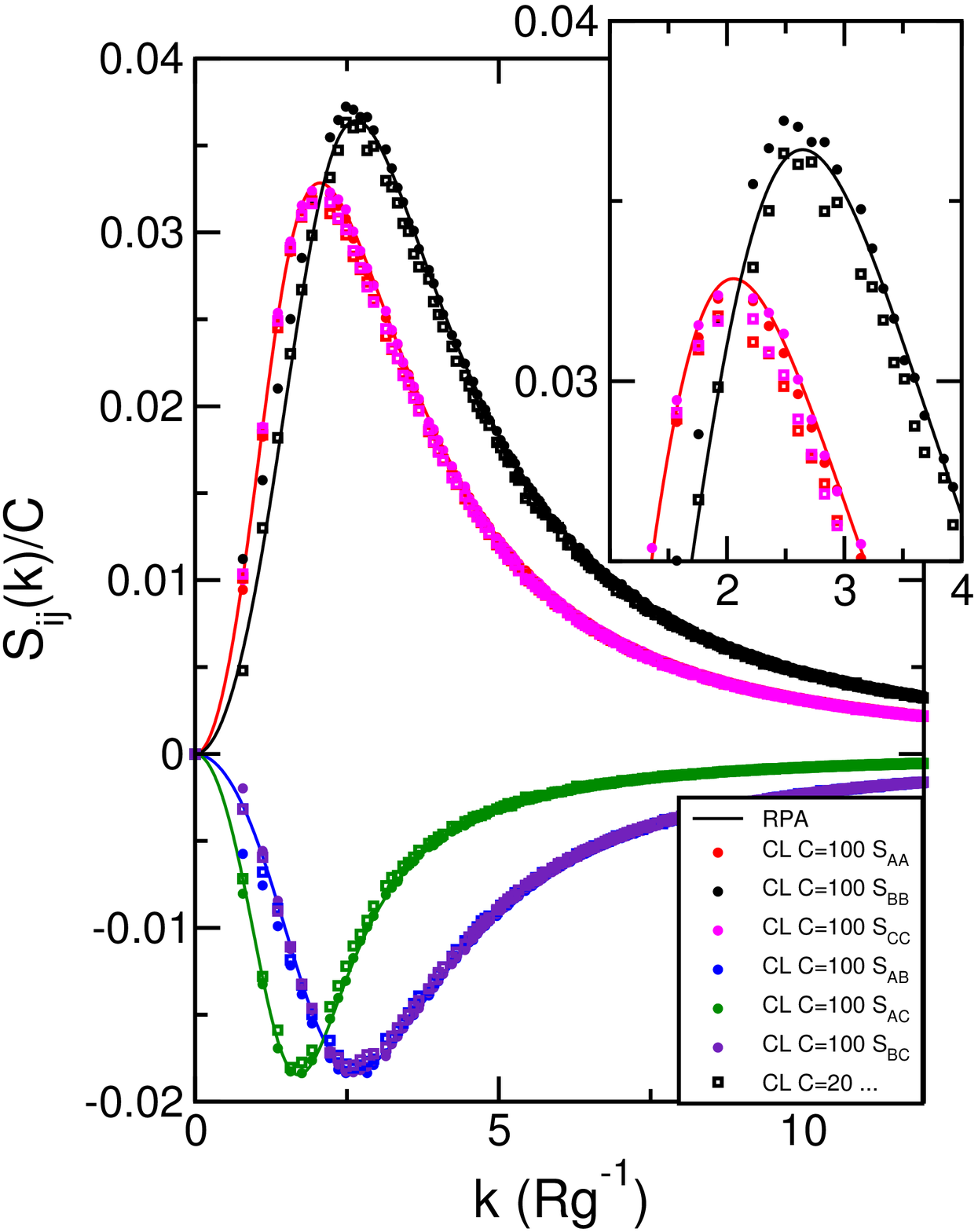}}
\caption{Six structure factor components of an ABC symmetric triblock melt in the disordered phase
with $f_B=0.6$, $\chi_{AB}N=\chi_{BC}N=10$, $\chi_{AC}N=14$. The simulation cell
is a cube of side $8$\,$R_g$ with a spatial collocation mesh of $32\times 32\times 32$.
High contour resolution is required to properly converge the high-$k$ decay of the structure factor; in
this case $400$ contour samples were used with a second-order operator splitting pseudospectral algorithm\cite{audus13}.
CL calculations were conducted at $C=100$ (weak fluctuations) and $C=20$ (moderately strong fluctuations).
Increasing fluctuation strength weakly dampens the amplitude of the dominant scattering peak (inset).
For comparison, the RPA structure factor is plotted (solid lines), which contains composition
fluctuations on the Gaussian level\cite{fredrickson06}.}
\label{fig:CLSK}
\end{center}
\end{figure}

Fig.\ \ref{fig:CLSK} shows the structure factor for an ABC triblock melt with $f_A = f_C = 0.2$ and $\chi_{AB}N = \chi_{BC}N = 10$, 
$\chi_{AC}N=14$ for both a normalized chain number density of $C=100$, corresponding to very high molecular weight and weak composition fluctuations, 
and $C=20$, smaller molecular weight with stronger fluctuations.
The disorder-phase structure factor contains a single peak resulting from short-length-scale correlations in the fluid. 
CL data are almost statistically indistinguishable from RPA curves for this system, which is sufficiently removed from the order-disorder transition. 

\paragraph*{Efficiency of weakly compressible CL simulations:}
\begin{figure}[H!t]
\begin{center}
{\includegraphics[width=\columnwidth]{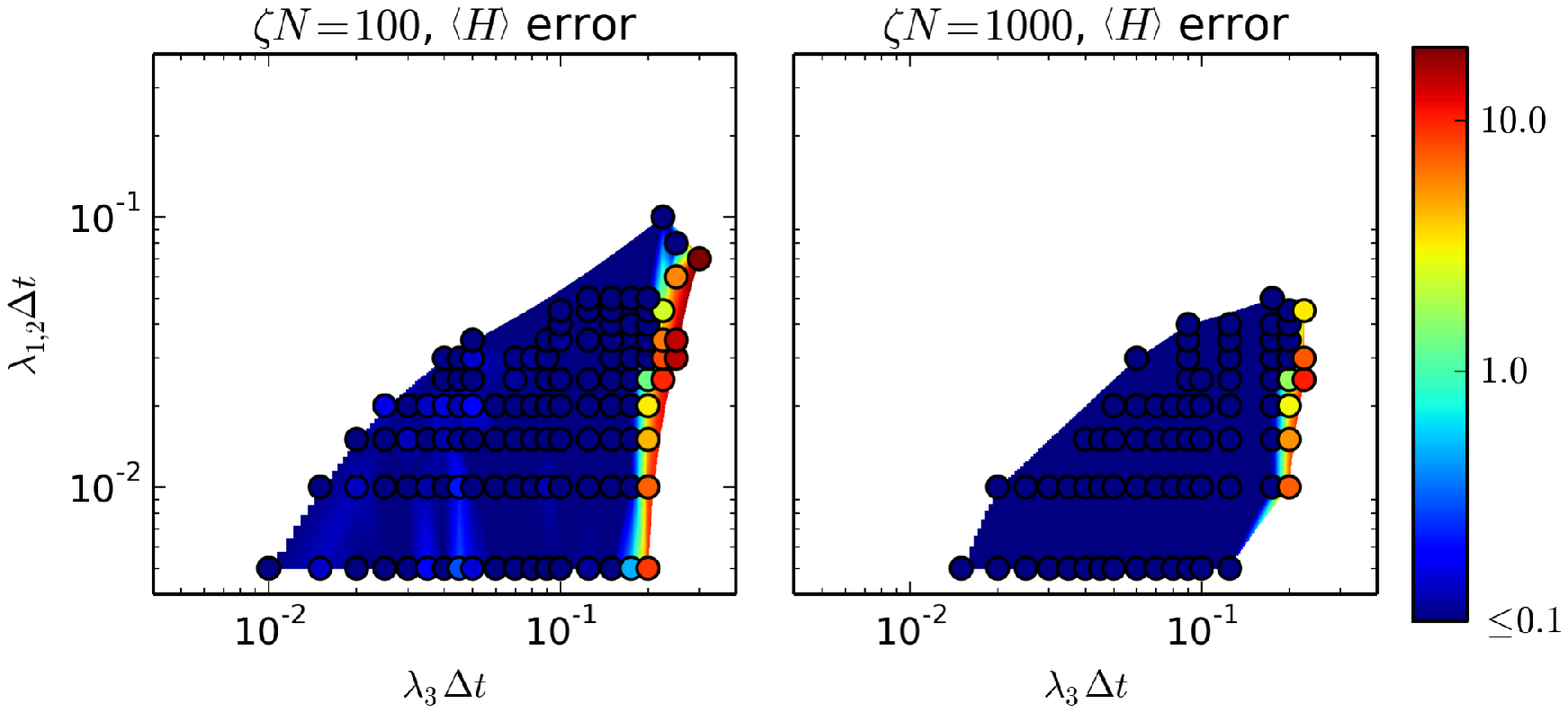}}
{\includegraphics[width=\columnwidth]{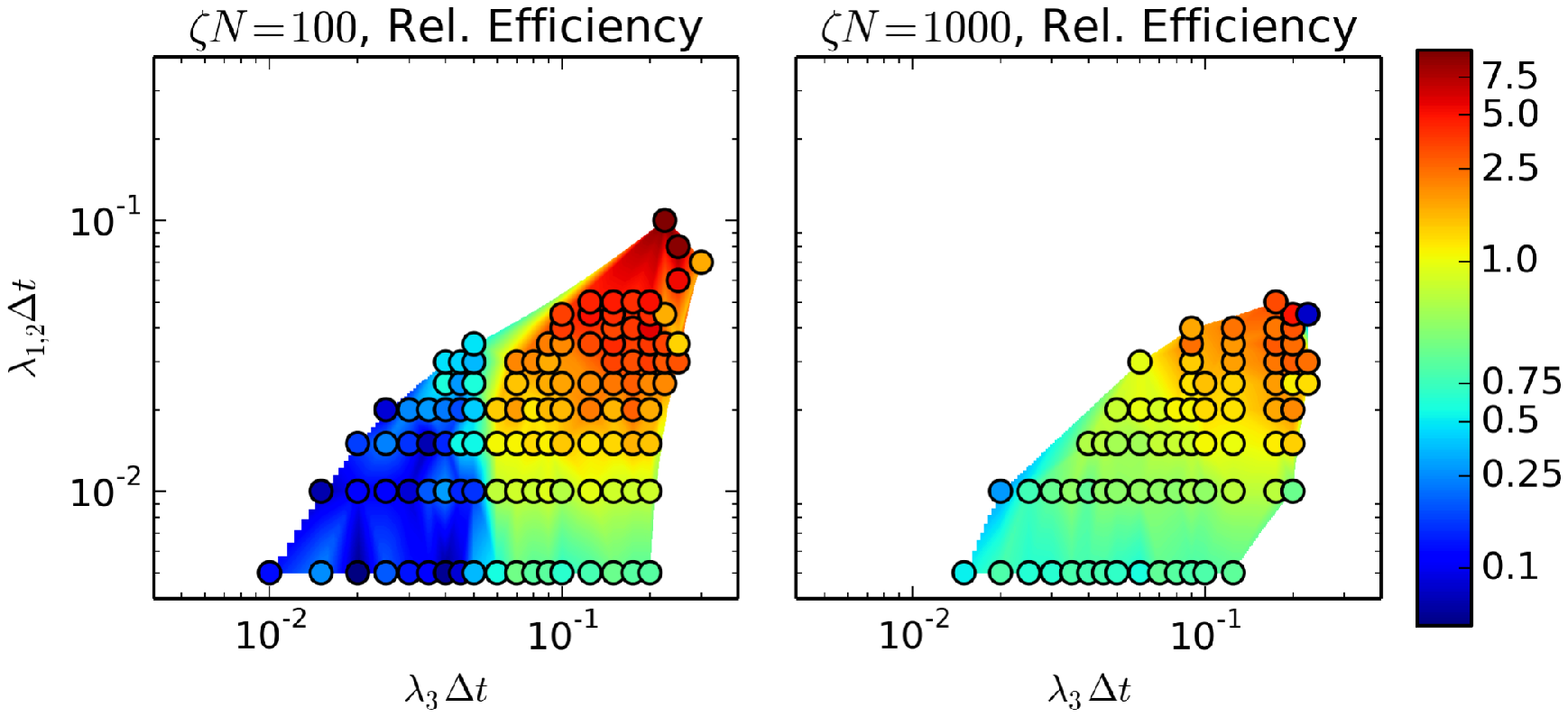}}
\caption{Complex Langevin simulations using the ETDPEC field updater for weakly compressible melts of ABC tiblock
copolymers in the alternating gyroid phase, with $f_A = f_C = 0.2$ and all $\chi N = 30$.
The upper panels show the error introduced in the estimate of $\left<H\right>$ as 
the mobility coefficients are varied, with reference  value taken from the small $\lambda_i$ limit with
the same compressibility. The lower panels show the relative efficiency, measured by
$\epsilon_{\zeta N}^2\left(\left\{\lambda_i\right\}\right)/\epsilon^2_{\infty,\mathrm{ref}}$, where $\epsilon$ is the standard error of the mean
of $\left<H\right>$ (here computed from $50,000$ CL time steps for each simulation). $\epsilon_{\infty,\mathrm{ref}}$ is a reference 
standard error of the mean taken from an incompressible simulation with optimal time step and $\lambda_i=1 \quad \forall i$. The efficiency
measures the relative length of simulation (number of samples) required to achieve a stochastic uncertainty of the estimate of $\left<H\right>$ that is
equal to the reference. Optimal choice of mobility coefficients can improve sampling efficiency by $\sim 10\times$--$100\times$.}
\label{fig:CLCompressible}
\end{center}
\end{figure}
Fig.\ \ref{fig:CLCompressible} presents data collected from complex Langevin simulations of a
weakly compressible ABC triblock melt in the alternating gyroid phase with $f_A = f_C = 0.2$ and
all binary $\chi N = 30$.
Simulation trajectories are more accurate for $\lambda_i \Delta t\rightarrow 0$, but the result is
a slower time evolution and longer correlation time so that a significant increase in the number of time steps 
is required to accurately evaluate thermally averaged operators.
The cost of simulations as mobility coefficients are varied can be evaluated using a relative efficiency
metric, $\epsilon_{\zeta N}^2\left(\left\{\lambda_i\right\}\right)/\epsilon^2_{\infty,\mathrm{ref}}$, where
$\epsilon$ is the standard error of the mean of $\left<H\right>$.
$\epsilon_{\infty,\mathrm{ref}}$ is a reference standard error taken from an incompressible simulation with
optimal time step ran for the same number of iterations.
An efficiency greater than $1.0$ implies that the simulation can be ran for proportionally fewer time steps
to achieve the same standard error as the reference.
In a compressible model with all binary $\chi N$ equal, the MSE eigenvalue spectrum is split as detailed in
Section \ref{sec:scft_compressible_results}.
Similarly grouping the mobility coefficients $\lambda_1 = \lambda_2 \neq \lambda_3$ in this case can significantly
improve both simulation stability \emph{and} efficiency, as shown in the lower panels of Fig.\ \ref{fig:CLCompressible}.

\section{Conclusions}
We have introduced a multi-species-exchange field-theoretic model for 
multi-species, multiblock polymers and blends with
arbitrary chain architectures.
We have demonstrated that the method can be used as a basis for stable
and accurate numerical simulations of the full field theory, including 
composition fluctuations, using the complex Langevin sampling scheme. 
Such fully fluctuating simulations were demonstrated to reproduce
RPA structure factors for high-molecular-weight 
triblock copolymer melts deep in the disordered phase.
In addition, mean-field (SCFT) solutions of the MSE model 
are readily available within the same framework and were shown to reproduce known phase boundaries
in an ABC triblock terpolymer melt

We have detailed the efficiency and accuracy of a variety of field update
algorithms for complex Langevin sampling, and explored the use of
asymmetric mobility coefficients to recover stability and efficiency in
weakly compressible melt models in both mean-field and non-mean-field simulations.

It is anticipated that the MSE method will be an important vehicle for the efficient study of fluctuation
phenomena, including beyond-mean-field corrections to phase diagrams of multi-species melts.
In addition, although a comparison has not been made in the present work, it is
conceivable that the MSE approach may perform better than traditional density-explicit species
methods for mean-field saddle-point searches, due in part to the
relaxation along normal modes of the interaction matrix.

We note that the ultra-violet sensitivity\cite{Villet2012,wang2010,fredrickson06} present in many polymer field theories remains
present in the simulations presented here.
However, a systematic strategy is available to regularize the field theory, which is entirely compatible
with the MSE scheme.\cite{Villet2012,wang2010}
We will explore this aspect in future work.

\subsection{Acknowledgements}
We acknowledge support from the Center for Scientific Computing at the
CNSI and MRL: an NSF MRSEC (DMR-1121053) and NSF CNS-0960316.
This work was partially supported by the CMMT program of the
National Science Foundation under Award No. DMR-1160895.
KTD was partially supported by the NSF DMREF program under Award DMR-1332842.
DD was partially supported by the Deutsche Forschungsgemeinschaft (DFG, German Research Foundation)
under DU 1100/1-1.

\appendix

\section{Exchange Mapping and Field Theory Transformation for Incompressible Multi-species Melt}
\label{sec:appendix_incomp}
We begin with the canonical partition function from Eqn.\ \ref{eqn:ce_incmelt_particlepartfn}.
Incompressibility can be enforced explicitly in the interaction term by replacing $\hat{\rho}_S = \rho_0 - \sum_{j=1}^{S-1} \hat{\rho}_j$.
Defining $\vc{P}^T = \left(\hat{\rho}_1,\ldots,\hat{\rho}_{S-1}\right)$, $\mat{X} = \left(\chi_{ij}N-2\chi_{iS}N\right)$, and $\vc{X}_S=\left(\chi_{iS}N\right)$ for $i,j\in\left[1,S-1\right]$, 
and using $\chi_{SS}=0$, we have
\begin{eqnarray}
  \beta U_1 & = & \frac{1}{2\rho_0N}\sum_{i,j=1}^{S-1}\rint \hat{\rho}_i\fnr \chi_{ij}N\hat{\rho}_j\fnr  \nonumber\\
            &+& \frac{1}{\rho_0N}\sum_{i=1}^{S-1}\rint \hat{\rho}_i\fnr \chi_{iS}N\left(\rho_0-\sum_{j=1}^{S-1}\hat{\rho}_j\fnr \right)\\
            & = & \frac{1}{2\rho_0N}\rint \left( \vc{P}^T\mat{X}\vc{P} + 2 \vc{P}^T\vc{X}_S\rho_0\right)\\
& = & \frac{1}{2\rho_0N}\rint  \left(\vc{P}+\mat{X}^{-1}\vc{X}_S\rho_0\right)^T\mat{X}\left(\vc{P}+\mat{X}^{-1}\vc{X}_S\rho_0\right) \nonumber\\
&-& \frac{1}{2\rho_0N}\rint  \vc{X}_S^T\mat{X}^{-1}\vc{X}_S\rho_0^2.
  \label{eqn:intterm_incompmelt_nondiag}
\end{eqnarray}
where the latter expression is obtained by completing the square.

We now diagonalize the $\mat{X}$ matrix using a similarity transform: $\mat{X}=\mat{O}\mat{D}\mat{O}^T$,
where $\mat{O}$ is an orthogonal matrix with columns equal to the eigenvectors of $\mat{X}$. 
$\mat{D}$ is a diagonal matrix containing the eigenvalues of $\mat{X}$, $\left\{d_i\right\}$.
Without loss of generality, we here assume that $\mat{X}$ has no zero eigenvalues, i.e., 
all $d_{i}$ are nonzero. 
The presence of any $d_{i}=0$ indicates a singular interaction matrix, pointing to
redundancies in the way the species are labeled. 
For example, a model with three species A, B, and C for which $\chi_{AB}=\chi_{AC}$ and $\chi_{BC}=0$
should be rewritten as a two-species model because B and C are indistinguishable.
All zero-eigenvalue modes can thus be deleted and the remaining modes renumbered.
Inserting the similarity transform into Eqn.~\ref{eqn:intterm_incompmelt_nondiag}, we find
\begin{eqnarray}
  \beta U_1 & = & \frac{1}{2\rho_0N}\rint  \sum_{i=1}^{S-1}d_i\left[\mat{O}^T\left(\vc{P}+\rho_0\mat{X}^{-1}\vc{X}_S\right)\right]_i^2 \nonumber\\
            &-& \frac{\rho_0}{2N}\rint  \sum_{i=1}^{S-1}\left[\mat{O}^T\vc{X}_S\right]_i^2 d_i^{-1},
  \label{eqn:incomp_uint_diagonalized}
\end{eqnarray}
where we have made use of the fact that $\mat{O}^{-1} = \mat{O}^T$, and $\left[\ldots\right]_i$ selects the $i$th component of a vector.
We now split Eqn.~\ref{eqn:incomp_uint_diagonalized} into two parts: $\beta U_1 = \beta\bar{U}_1 + \beta\bar{U}_\mathrm{ref}$,
where $\beta\bar{U}_\mathrm{ref}$ depends only on the number of species and their mutual $\chi$ interactions. 
With convenient non-dimensionalizations:
\begin{equation}
  \beta\bar{U}_\mathrm{ref} = -\frac{CV}{2}\sum_{i=1}^{S-1}\left(\sum_{j=1}^{S-1} O_{ji}\chi_{jS}N\right)^2d_i^{-1},
\end{equation}
which depends only on the $\left\{ \chi N\right\}$ parameters together with the dimensionless chain density, $C$, and the cell volume, $V$.
This constant term, which can be precomputed, enters SCFT and CL calculations only as a free-energy shift; it does
not depend on $n_p$ for any of the molecular constituents, and consequently does not affect the
chemical potential references. 
The remaining part of the interaction term can be written as
\begin{equation}
  \beta \bar{U}_1 = \frac{1}{2\rho_0N}\rint \sum_{i=1}^{S-1} d_i \sigma_i^2\fnr ,
  \label{eqn:quadraticform}
\end{equation}
where 
$\sigma_i\fnr  = \left[\mat{O}^T\left(\vc{P}\fnr +\rho_0\mat{X}^{-1}\vc{X}_S\right)\right]_i$ 
are linear combinations of the original 
microscopic density operators, $\hat{\rho}_i\fnr$, with 
$\mat{\chi} N$-dependent constant shifts originating from completing the square.
The purely quadratic form of Eqn.\ \ref{eqn:quadraticform} permits a direct decoupling of interactions through
Hubbard-Stratonovich transformation.
For this task we make use of the following Gaussian functional integral (GFI) identity 
(Eqns. C.27 \& C.28 of Ref.\ \onlinecite{fredrickson06}):
\begin{widetext}
\begin{equation}
  \exp\left(\frac{\gamma^2}{2}\rint  \int d\vect{r}^\prime J\fnr A^{-1}\left(\vect{r},\vect{r}^\prime\right)J\left(\vect{r}^\prime\right)\right) = \frac{\int \mathcal{D}f \exp\left(-\frac{1}{2}\rint \int d\vect{r}^\prime f\fnr A\left(\vect{r},\vect{r}^\prime\right)f\left(\vect{r}^\prime\right) - \gamma\rint J\fnr f\fnr \right)}{\int \mathcal{D}f \exp\left(-\frac{1}{2}\rint \int d\vect{r}^\prime f\fnr A\left(\vect{r},\vect{r}^\prime\right)f\left(\vect{r}^\prime\right)\right)},
  \label{eqn:HSDefinition}
\end{equation}
where $\gamma = \pm i$ or $\pm 1$.
In this definition of the Hubbard-Stratonovich transformation, $\fnint f$ is a functional integral over real-valued configurations of the field $f\fnr$.
Let $A^{-1}\left(\vect{r},\vect{r}^\prime\right) = \frac{\left|d_i\right|}{\rho_0N}\delta\left(\vect{r}-\vect{r}^\prime\right)$, then
\begin{eqnarray}
  e^{-\beta\bar{U}_1} & = & \prod_{i=1}^{S-1} \exp\left(\frac{\left|d_i\right|\gamma_i^2}{2\rho_0N}\rint  \sigma_i^2\fnr \right)\\
                      &=& \prod_{i=1}^{S-1}\left(D_i^{-1} \int \mathcal{D}w_i \exp\left(-\frac{\rho_0N}{2\left|d_i\right|}\rint  w_i^2\fnr  - \gamma_i\rint  w_i\fnr \sigma_i\fnr \right)\right),
\end{eqnarray}
where we have used $\gamma_i = 1$ for $d_i<0$ and $\gamma_i=i$ otherwise. 
$w_i\fnr$ is an auxiliary chemical potential field that is conjugate to $\sigma_i\fnr$ -- the exchange-mapped linear combination of species microscopic density operators.
$D_i$ is a Gaussian functional integral denominator that contributes to the overall measure of the partition function:
\begin{equation}
 D_i = \int \mathcal{D}w_i \exp\left(-\frac{\rho_0N}{2\left|d_i\right|}\rint  w_i^2\fnr  \right).
\end{equation}
Hence,
\begin{equation}
  \mathcal{Z}_c = \mathcal{Z}_\mathrm{ig} \left(\prod_{k=1}^{S-1}D^{-1}_k\right)\int\prod_{i=1}^n\mathcal{D}\vect{r}_i\int\prod_{j=1}^{S-1}\mathcal{D}w_j \exp\left[-\beta U_0 - \sum_i\left(\frac{\rho_0N}{2\left|d_i\right|}\rint  w_i^2\fnr  + \gamma_i \rint  w_i\fnr \sigma_i\fnr \right)\right]\delta\left[\hat{\rho}_+\fnr -\rho_0\right],
\end{equation}
where $\hat{\rho}_+\fnr  = \sum_i^S\hat{\rho}_i\fnr $ and $\mathcal{Z}_\mathrm{ig} = \exp\left(-\beta \bar{U}_\mathrm{ref}\right)\left(\prod_{p=1}^P\lambda_T^{3n_pN_p}n_p!\right)^{-1}$.

We now transform the functional delta to exponential form\cite{fredrickson06}. 
This leads to
\begin{eqnarray}
  \mathcal{Z}_c = \mathcal{Z}_\mathrm{ig}^\prime \int\prod_{i=1}^n\mathcal{D}\vect{r}_i\,e^{-\beta U_0}\int\prod_{j=1}^{S-1}\mathcal{D}w_j\int \mathcal{D}w_+ &\exp&\left[-\sum_{i=1}^{S-1}\left(\frac{\rho_0N}{2\left|d_i\right|}\rint  w_i^2\fnr  +\gamma_i \rint  w_i\fnr \sigma_i\fnr \right)\right.\nonumber\\
                                                                                                                                                    &-&\left.i\rint w_+\fnr \left(\hat{\rho}_+\fnr -\rho_0\right)\right],
\end{eqnarray}
with
\begin{equation}
  \mathcal{Z}_\mathrm{ig}^\prime = \frac{1}{\prod_{p=1}^P\lambda_T^{3n_pN_p} n_p!}e^{-\beta\bar{U}_\mathrm{ref}}\left(\int\prod_{i=1}^{S-1}\mathcal{D}w_i\,e^{-\sum_{i=1}^{S-1}\frac{\rho_0N}{2\left|d_i\right|}\rint  w_i^2\fnr }\right)^{-1}\left(\int \mathcal{D}w\, e^{-\frac{1}{2}\rint w^2\fnr }\right)^{-2}.
\end{equation}
Upon re-inserting the full expression for $\sigma_i$, we find
\begin{eqnarray}
  \mathcal{Z}_c &=& \mathcal{Z}_\mathrm{ig}^\prime \int\prod_{j=1}^{S-1}\mathcal{D}w_j\fnint w_+\, e^{-\sum_{i=1}^{S-1}\frac{\rho_0N}{2\left|d_i\right|}\rint  w_i^2\fnr  + i\rint w_+\fnr \rho_0}\nonumber\\
              &\times& \int\prod_{i=1}^n\mathcal{D}\vect{r}_i\,e^{-\beta U_0 - \sum_{i,j=1}^{S-1}\gamma_i O_{ji}\rint w_i\fnr \left(\hat{\rho}_j\fnr +\rho_0\sum_{k=1}^{S-1} X_{jk}^{-1}\chi_{kS}N\right) - i\rint  w_+\fnr \sum_{i=1}^{S}\hat{\rho}_i\fnr }\\
  \Rightarrow \mathcal{Z}_c &=& \mathcal{Z}_\mathrm{ig}^\prime \int\prod_{j=1}^{S-1}\mathcal{D}w_j\fnint w_+\, e^{-\sum_{i=1}^{S-1}\frac{\rho_0N}{2\left|d_i\right|}\rint  w_i^2\fnr  - \rho_0\sum_{i,j=1}^{S-1} \gamma_i O_{ji}\sum_{k=1}^{S-1}X_{jk}^{-1}\chi_{kS}N\rint w_i\fnr + i\rint w_+\fnr \rho_0}\nonumber\\
              &\times& \int\prod_{i=1}^n\mathcal{D}\vect{r}_i\,e^{-\beta U_0 - \sum_{i,j=1}^{S-1}\gamma_i O_{ji}\rint w_i\fnr \hat{\rho}_j\fnr  - i\rint  w_+\fnr \sum_{i=1}^{S}\hat{\rho}_i\fnr }.
  \label{eqn:ZcInt}
\end{eqnarray}
\end{widetext}
The final part of Eqn.\ \ref{eqn:ZcInt} is a partition function of non-interacting chains subject to external fields 
$\left\{w_i\fnr\right\}$. 
Note that $\psi_j\fnr := \sum_i O_{ji}\gamma_i w_i\fnr+iw_+\fnr$ is conjugate to $\hat{\rho}_j\fnr$, and as such
$\psi_j$ can be defined as ``species'' chemical potential fields.

We now make the following transformation to the exchange-mapped auxiliary 
fields $\left\{w_{i,+}\right\}$:
$\mu_i\fnr  = \gamma_i w_i\fnr  N$, $\mu_+\fnr  = iw_+\fnr N$.
This mapping includes a Wick rotation for pressure-like fields -- those with $d_i>0$ and $w_+$.
We do this in all terms in $\mathcal{Z}_c$, including the 
Gaussian functional integral denominators in 
$\mathcal{Z}_\mathrm{ig}^\prime$, so that the implied Jacobians cancel regardless of 
discretization strategy.
Note that Wick rotations of the subset of the fields with pressure character 
changes their integration domain so that $\int\mathcal{D}\mu$ is over 
purely imaginary fields for that subset.
%
With this mapping, the partition function becomes
\begin{widetext}
\begin{eqnarray}
  \mathcal{Z}_c &=& \mathcal{Z}_{ig}^\prime \int\prod_{j=1}^{S-1}\mathcal{D}\mu_j\fnint \mu_+\, e^{\sum_{i=1}^{S-1}\frac{C}{2d_i}\rint  \mu_i^2\fnr  - C\sum_{i,j=1}^{S-1}  O_{ji}\sum_{k=1}^{S-1}X_{jk}^{-1}\chi_{kS}N\rint \mu_i\fnr + C\rint \mu_+\fnr }\nonumber\\
                &\times& \int\prod_{i=1}^n\mathcal{D}\vect{r}_i\,e^{-\beta U_0 - \sum_{i,j=1}^{S} A_{ij}\rint \mu_j\fnr \hat{\rho}_i\fnr /N},
\label{eqn:beforeparticleintegrate}
\end{eqnarray}
\end{widetext}
where $C=\rho_0R_g^3/N$ is a dimensionless polymer chain number density, and we have non-dimensionalized all lengths with respect
to the radius of gyration of a reference ideal polymer chain, $R_g = b\left(N/6\right)^{1/2}$ with a reference
statistical segment length $b$ and degree of polymerization $N$.
We have made use of the fact that $\left|d_i\right|\gamma_i^2 = -d_i \enskip \forall i$.
The factor $1/N$ in the latter terms can be absorbed by rescaling the 
contour variable to $\left[0,\alpha_p\right]$ in $\hat{\rho}$.
We have defined the matrix $A_{ij}$ that transforms from 
exchange fields, $\left(\left\{\mu_i\fnr\right\}, \mu_+\fnr\right)$, to the 
species fields $\left(\psi_A\fnr N, \ldots, \psi_S\fnr N\right)$ that are directly
conjugate to $\hat{\rho}_i\fnr$, $i\in\left[1,S\right]$:
\begin{equation}
A_{ij} = \left(
  \begin{array}{cc}
    O_{ij} & e_i\\
   0 & 1
 \end{array}
 \right),
\end{equation}
where $\vc{e}$ is a column vector containing ones with $S-1$ entries.
Note that $\mu_+\fnr = \psi_S\fnr N$. 
Due to the orthogonal nature of $\mat{O}$, the inverse of $A$ is readily identified to be
\begin{equation}
  A^{-1}_{ij} = \left(\begin{array}{cc}
  O_{ji} & X_i\\
  0 & 1
  \end{array}\right),
\end{equation}
where $X_i$ is a (column) vector $X_i = -\sum_{k=1}^{S-1}O_{ki}$ with $S-1$ entries.
Though $A^{-1}_{ij}$ is never needed during the course of a field theoretic simulation, in which the fundamental degrees of freedom
are the fields $\left\{\mu_i\right\}$, $\mu_+$ entering the functional integrals, having this transformation can be useful for seeding a
calculation using the species fields.
Initialization with species fields is preferred, because then field definitions are 
independent of the exchange-mapping strategy (e.g., elimination of a species other than $S$ when applying the
incompressibility constraint).

Isolating the single-chain part of Eqn.~\ref{eqn:beforeparticleintegrate}, we have
\begin{widetext}
\begin{equation}
  \int \prod_{p=1}^P\prod_{l=1}^{n_p} \mathcal{D}\vect{r}_l^p \exp\left(-\sum_{p=1}^{P}\sum_{l=1}^{n_p}\int_0^{\alpha_p}ds\,\frac{3N}{2b\left(s\right)^2}\left|\frac{d\vect{r}_l^p\left(s\right)}{ds}\right|-\sum_p^P\sum_l^{n_p}\sum_i^S\int_{s\in i}ds\,N\psi_i\left(\vect{r}_l^p\left(s\right)\right)\right),
\end{equation}
\end{widetext}
where $i$ indexes chemical species, $p$ indexes chain types, and $l$ indexes members of each type.
The contour variable $s$ has been rescaled by $N$.
In this expression, the microscopic density operator has been substituted and the $\delta$ functions over particle coordinates
evaluated, $\vc{\psi}=\mat{A}\vc{\mu}$
Evaluating $\sum_i$ and expanding the $\exp\left(\ldots\right)$ leads to
\begin{widetext}
\begin{equation}
  \prod_{p=1}^P\prod_{l=1}^{n_p}\left[\int \mathcal{D}\vect{r}_l^p \exp\left(-\int_0^{\alpha_p}ds\,\frac{3N}{2b\left(s\right)^2}\left|\frac{d\vect{r}_l^p\left(s\right)}{ds}\right|-\int_0^{\alpha_p}ds\,N\psi_p\left(\vect{r}_l^p\left(s\right);s\right)\right)\right]:=\prod_{p=1}^P\left(V g_{N_p} Q_p\left[\mat{A}\vc{\mu}\right]\right)^{n_p}
\end{equation}
where $Vg_{N_p}$ is the reference single-chain partition function of an ideal polymer 
in free space of volume $V$ (Ref.\ \onlinecite{fredrickson06}), and
$\psi_p\left(\vect{r}\left(s\right);s\right):=\psi_i\left(\vect{r}\left(s\right)\right)$ 
for $s$ in the $i$ block (i.e., $\psi_p$ has a parametric dependence on $s$, 
switching between the different species fields depending on the 
statistical segment on which it is acting).

Now use the following relation: $\sum_j O_{ji}X^{-1}_{jk} = \mat{O}^T\mat{X}^{-1} = \left(\mat{X}\mat{O}\right)^{-1} = \mat{D}^{-1}\mat{O}^T$ to give $C\sum_{i,j,k=1}^{S-1} O_{ji}X^{-1}_{jk}\chi_{kS} = C\sum_{i,j=1}^{S-1}\left(d_i\right)^{-1}O_{ji}\chi_{jS}N$.
Then the partition function becomes
\begin{eqnarray}
  \mathcal{Z}_c &=& \mathcal{Z}_0 \fnint \mu_1\ldots\fnint \mu_{S-1}\fnint \mu_+\, e^{-H\left[\left\{\mu_i\right\},\mu_+\right]}\\
  \label{eqn:H_incomp_mse}
  H\left[\left\{\mu_i\right\},\mu_+\right] &=& -\sum_{i=1}^{S-1}\frac{C}{2d_i}\rint  \mu_i^2\fnr  + \sum_{i,j=1}^{S-1}\frac{C}{d_i}  O_{ji}\chi_{jS}N\rint \mu_i\fnr - C\rint \mu_+\fnr  - \sum_{p=1}^P n_p \ln Q_p\left[\mat{A}\vc{\mu}\right],
\end{eqnarray}
\end{widetext}
where $\mathcal{Z}_0 = \mathcal{Z}_{ig}^\prime V^n \prod_{p=1}^P g_{N_p}^{n_p}$.
(i.e., $\mathcal{Z}_0$ now contains all terms required for an ideal gas of polymers chains,
together with the Gaussian denominators introduced during Hubbard-Stratonovich transformations).
It is convenient to always work with volume fractions instead of explicit particle/molecule numbers.
The $n_p$ of the final term can be replaced as follows:
\begin{eqnarray}
  \rho_0 &=& \sum_p \frac{n_pN_p}{V} = CN \\
  \Rightarrow CV &= &\sum_p n_p \alpha_p\\
  \phi_p &:= & \frac{n_p N_p}{\sum_q n_q N_q} \equiv \frac{n_p \alpha_p}{\sum_q n_q \alpha_q}\\
  \Rightarrow n_p & = & \frac{CV\phi_p}{\alpha_p},
\end{eqnarray}
where $\phi_p$ is the chain volume fraction (i.e., the fraction of the total count of monomers originating on chains of type $p$),
and $\alpha_p$ is the chain length relative to the reference $N$. 
Similar derivations for branched polymers result in the $\alpha_p$ denominator being replaced by a sum of sub-chain lengths relative to $N$.

The normalized single-chain partition function for a linear chain in a collection of external fields can be written
\begin{equation}
  Q_p\left[\mat{A}\vc{\mu}\right] = \frac{1}{V}\rint  q_p\left(\vect{r},\alpha_p\right),
\end{equation}
where $q_p$ is computed through
\begin{equation}
  \partial_s q\left(\vect{r},s\right) = \frac{b\left(s\right)^2}{b^2} \nabla^2 q\left(\vect{r},s\right) - N\psi_p\left(\vect{r};s\right)q\left(\vect{r},s\right),
\end{equation}
with $s\in\left[0,\alpha_p\right]$, and $q\left(\vect{r},0\right) = 1$ is the initial condition for untethered, unconstrained polymer chains.
All lengths are scaled to units of $R_g = b\left(N/6\right)^{1/2}$.

\section{Exchange Mapping and Field Theory Transformation for Compressible Multi-species Melt}
\label{sec:appendix_comp}
In the case of a weakly compressible melt, we do not eliminate one of the species in favor of the total polymer density ---  all spatially
inhomogeneous microscopic density operators must be retained. 
The strategy here involves completing the square in the exponent, diagonalizing
the resulting coupling matrix, and making a Hubbard-Stratonovich transformation of all
normal-mode density operators.
We will use the symbol $\vc{e}=\left(1,\ldots,1\right)^T$, a rank-one vector with $S$ elements, and \mat{\openone} an $S\times S$ matrix with all entries equal $1$.
Consider $\beta U_\mathrm{int} = \beta U_1 + \beta U_2$ from Eqn.\ \ref{eqn:ce_hcmelt_particlepartfn}:
\begin{widetext}
\begin{eqnarray}
  \beta U_\mathrm{int} & = & \frac{1}{2\rho_0}\rint  \left[\vc{\rho}^T\fnr \mat{\chi}\vc{\rho}\fnr  + \zeta\left(\vc{\rho}^T\fnr \vc{e}-\rho_0\right)^2\right]\\
                       & = & \frac{1}{2\rho_0}\rint  \left[\vc{\rho}^T\fnr \mat{\chi}\vc{\rho}\fnr  + \zeta\vc{\rho}^T\fnr \vc{\rho}\fnr  - 2\zeta\rho_0\vc{\rho}^T\fnr \vc{e}+\zeta\rho_0^2\right]\\
                       & = & \frac{1}{2\rho_0 N}\rint  \left[\left(\vc{\rho}\fnr +\vc{\nu}\right)^T\left(\mat{\chi}N+\zeta N\right)\left(\vc{\rho}\fnr  + \vc{\nu}\right) - \vc{\nu}^T\left(\mat{\chi}N + \zeta N\right)\vc{\nu}+\zeta N\rho_0^2\right]
\end{eqnarray}
\end{widetext}
where the last step comes from completing the square, which demands that $\vc{\nu} = -\rho_0\zeta N\left(\mat{\chi}N+\zeta N\right)^{-1}\vc{e}$.
The symbols $\vc{\rho}\fnr $, $\mat{\chi}$ and $\rho_0$ are as defined in Appendix \ref{sec:appendix_incomp}.
One option to proceed from here is to directly diagonalize the matrix $\left(\mat{\chi}N+\zeta N\right)$ to decompose $\exp\left(-\beta U_\mathrm{int}\right)$ 
into normal modes.
However, the large difference in the common magnitude of $\left\{\chi_{ij} N\right\}$ versus that of $\zeta N$ can cause numerical problems.
The matrix becomes singular for $\zeta N$ large (approaching the incompressible limit) which makes the inversion involved in $\vc{\nu}$ problematic.
Likewise, extracting a factor of $\zeta N$ from the matrix results in $\chi N / \zeta N$, which approaches zero (numerical under-run) when a strong 
compressibility penalty is imposed, also resulting numerically in a singular matrix. 
For typical magnitudes of the matrix entries, extracting a factor of 
$\left(\zeta N\right)^{-\frac{1}{2}}$ tends to make the eigenvalues close to order $1$.
Then
\begin{eqnarray}
  \beta U_\mathrm{int} &=& \frac{\left(\zeta N\right)^\frac{1}{2}}{2\rho_0 N}\rint \left[\left(\vc{\rho}\fnr +\vc{\nu}\right)^T\mat{X}\left(\vc{\rho}\fnr  + \vc{\nu}\right) - \vc{\nu}^T\mat{X}\vc{\nu}\right]\nonumber\\
                       &+&\frac{\rho_0V\zeta N}{2N},
\end{eqnarray}
with $\mat{X} = \left(\mat{\chi}N\left(\zeta N\right)^{-\frac{1}{2}} + \left(\zeta N\right)^\frac{1}{2}\mat{\openone}\right)$, and 
$\vc{\nu} = -\rho_0 \left(\zeta N\right)^\frac{1}{2}\mat{X}^{-1}\vc{e}$.
We now diagonalize the matrix using $\mat{X} =\mat{O}\mat{D}\mat{O}^T$.
\begin{eqnarray}
  \beta U_\mathrm{int} & = & \frac{\left(\zeta N\right)^\frac{1}{2}}{2\rho_0 N}\rint  \left[\left(\vc{\rho}\fnr +\vc{\nu}\right)^T\mat{O}\mat{D}\mat{O}^T\left(\vc{\rho}\fnr + \vc{\nu}\right) \right.\nonumber\\
                       &-& \left. \vc{\nu}^T\mat{X}\vc{\nu}+\rho_0^2\left(\zeta N\right)^\frac{1}{2}\right]\\
                       & = & \frac{\left(\zeta N\right)^\frac{1}{2}}{2\rho_0 N}\rint  \left[\sum_{i=1}^S d_i \sigma_i^2\fnr  + K\right]
\end{eqnarray}
where $\sigma_i\fnr  = \sum_{j=1}^S O_{ji}\left(\hat{\rho}_j\fnr  + \nu_j\right)$ and $K = \rho_0^2\left(\zeta N\right)^\frac{1}{2} - \rho_0^2 \zeta N\sum_{i,j}^SX^{-1}_{ij}$, 
where the last term comes from $\vc{\nu}^T\mat{X}\vc{\nu} = \rho_0^2 \vc{e}^T\left(\zeta N\right)^\frac{1}{2}\left(\mat{X}^{-1}\right)^T\mat{X}\mat{X}^{-1}\left(\zeta N\right)^\frac{1}{2}\vc{e} = \rho_0^2\zeta N\sum_{i,j}X^{-1}_{ij}$.
Now we split the interaction energy into a constant reference shift and a remainder that depends on the system configuration.
\begin{eqnarray}
  \beta U_\mathrm{ref}&=&\frac{CV}{2\left(\zeta N\right)^{-1}}\left(1-\frac{1}{\left(\zeta N\right)^{-\frac{1}{2}}}\sum_{i,j=1}^SX_{ij}^{-1}\right)\\
  &=&\frac{CV}{2\left(\zeta N\right)^{-1}}\left(1-\frac{1}{\left(\zeta N\right)^{-\frac{1}{2}}}\sum_{i=1}^S d_i^{-1}\left(\sum_{j=1}^S O_{ji}\right)^2\right).\nonumber\\
\end{eqnarray}
The Hubbard-Stratonovich transformations introduced in Equation \ref{eqn:HSDefinition} can be used with the positive-definite interaction potential 
$A_i^{-1} \left(\vect{r},\vect{r}^\prime\right) = \frac{\left|d_i\right|\left(\zeta N\right)^\frac{1}{2}}{\rho_0N}\delta\left(\vect{r}-\vect{r}^\prime\right)$.
$-d_i = \left|d_i\right|\gamma_i^2$ with $\gamma_i = \pm i$ if $d_i>0$, $\gamma_i = \pm 1$ otherwise. 
\begin{widetext}
\begin{eqnarray}
  e^{-\beta U_\mathrm{int}} & = & e^{-\beta U_\mathrm{ref}}\prod_{i=1}^S\left[\exp\left(\gamma_i^2\frac{\left|d_i\right|\left(\zeta N\right)^\frac{1}{2}}{2\rho_0N}\rint \sigma_i^2\fnr \right)\right]\\ 
                            & = & e^{-\beta U_\mathrm{ref}}\prod_{i=1}^S\left[D_i^{-1}\int \mathcal{D}w_i \exp\left(-\frac{\rho_0N}{2\left|d_i\right|\left(\zeta N\right)^\frac{1}{2}}\rint  w_i^2\fnr -\gamma_i\rint  w_i\fnr \sigma_i\fnr \right)\right]
\end{eqnarray}
The Gaussian functional integral denominator is
\begin{equation}
  D_i = \int \mathcal{D} w \exp\left(-\frac{C}{2\left|d_i\right|\left(\zeta N\right)^\frac{1}{2}}\rint N^2 w^2\fnr \right)
\end{equation}
Now substituting $\vc{\sigma}\fnr $ and collecting terms that depend on particle coordinates:
\begin{eqnarray}
  e^{-\beta U_\mathrm{int}} & = & e^{-\beta U_\mathrm{ref}}\fnint  \left\{w_i\right\} \exp\left(-\sum_i^S\frac{\rho_0 N}{2\left|d_i\right|\left(\zeta N\right)^\frac{1}{2}}\rint  w_i^2\fnr -\sum_i^S\gamma_i\rint  w_i\fnr \left[\mat{O}^T\vc{\nu}\right]_i\right)\nonumber\\
                            &\times& \exp\left(-\sum_{i=1}^S\gamma_i\rint  w_i\fnr \left[\mat{O}^T\vc{\rho}\fnr \right]\right)\prod_{i=1}^SD_i^{-1},\\
\end{eqnarray}
\end{widetext}
Since $\mat{O}$ is orthogonal, $\mat{X} = \mat{O}\mat{D}\mat{O}^T$, and $\vc{\nu} = -\rho_0\left(\zeta N\right)^\frac{1}{2}\mat{X}^{-1}\vc{e}$, 
then $\mat{O}^T\vc{\nu} = -\rho_0\left(\zeta N\right)^\frac{1}{2}\left(\mat{O}\mat{D}\right)^{-1}\vc{e}$ and
$\left[\left(\mat{O}\mat{D}\right)\vc{e}\right]_i=\sum_j d_i^{-1} O_{ji}$.
We now make the replacement $\gamma_iNw_i\fnr  = \mu_i\fnr $, which includes a Wick-rotation for repulsive modes and 
a factor to scale out the reference polymer chain length.
\begin{widetext}
\begin{eqnarray}
  e^{-\beta U_\mathrm{int}} & = & e^{-\beta U_\mathrm{ref}}\fnint  \left\{\mu_i\right\} \exp\left(\sum_i^S\frac{C\left(\zeta N\right)^{-\frac{1}{2}}}{2d_i}\rint  \mu_i^2\fnr +\frac{C}{\left(\zeta N\right)^{-\frac{1}{2}}}\sum_{i,j=1}^S\rint  \frac{O_{ji}\mu_i\fnr}{d_i}\right)\nonumber\\
                            &\times& \exp\left(-\sum_{i=1}^S\rint  \mu_i\fnr \sum_{j=1}^S O_{ji}\hat{\rho}_j\fnr /N\right)\prod_{i=1}^SD_i^{-1},
\end{eqnarray}

The decoupled interaction potential is inserted into the partition function such that
\begin{eqnarray}
  \mathcal{Z}_c & = & \mathcal{Z}_\mathrm{ig} e^{-\beta U_\mathrm{ref}}\fnint\left\{\mu_i\right\}\,
  \exp\left(\sum_{i=1}^S\frac{C\left(\zeta N\right)^{-\frac{1}{2}}}{2d_i}\rint \mu_i^2\fnr +\frac{C}{\left(\zeta N\right)^{-\frac{1}{2}}}\sum_{i,j=1}^S\rint \frac{O_{ji}\mu_i\fnr}{d_i}\right)\nonumber\\
                &\times & \fnint\left\{\vect{r}_i\right\} \exp\left(-\beta U_0 - \sum_{i=1}^S\rint \psi_i\fnr\hat{\rho}_i\fnr/N\right)
\end{eqnarray}
where $\psi_i\fnr N := \sum_j O_{ij} \mu_j\fnr$, i.e., defining a species field that is conjugate to $\hat{\rho}_i\fnr$.
$\mathcal{Z}_\mathrm{ig} = \left(\prod_{p=1}^P\lambda_T^{3n_pN_p}n_p!\right)^{-1}\prod_{i=1}^S D_i^{-1}$.

Finally,
\begin{equation}
  \mathcal{Z}_c = \mathcal{Z}_0 \fnint \left\{\mu_i\right\} e^{-H\left[\left\{\mu_i\right\}\right]},
\end{equation}
where
\begin{eqnarray}
  \mathcal{Z}_0 & = & \left[\prod_{p=1}^P\frac{V^{n_p}g_{N_p}^{n_p}}{\lambda_T^{3n_pN_p}n_p!}\right]e^{-\beta U_\mathrm{ref}}\prod_{i=1}^SD_i^{-1}\\
  H\left[\left\{\mu_i\right\}\right] & = & -\sum_{i=1}^S \frac{C\left(\zeta N\right)^{-\frac{1}{2}}}{2d_i}\rint \mu_i^2\fnr-\frac{C}{\left(\zeta N\right)^{-\frac{1}{2}}}\sum_{i,j=1}^S\rint\frac{O_{ji}\mu_i\fnr}{d_i} - CV\sum_{p=1}^P \frac{\phi_p}{\alpha_p}\ln Q_p\left[\mat{O}\vc{\mu}\right]\\
  Q_p\left[\left\{\psi_i\right\}\right] & = & \left(Vg_{N_p}\right)^{-1}\left[\fnint\vect{r}^p\exp\left(-\frac{3N}{2b^2}\int_0^{\alpha_p}ds\,\left|\frac{d\vect{r}^p}{ds}\right|^2-\sum_{i=1}^S\int_{s\in i}ds\,\psi_i\left(\vect{r}^p\left(s\right)\right)\right)\right]\\
                                        & = & \frac{1}{V}\rint q_p\left(\vect{r},s=\alpha_p\right)
\end{eqnarray}
\end{widetext}
Note that the compressibility penalty appears as $\left(\zeta N\right)^{-1/2}$ everywhere, including in the 
$\mat{X}$ matrix. 
We use $\left(\zeta N\right)^{-1}$ as a convenient control parameter that recovers the incompressible limit at $0$.

\bibliography{mse_refs}
\end{document}